\documentclass[12pt]{amsart}
\usepackage{txfonts}      %{article} was 12pt latex e
\usepackage{amssymb}
\usepackage[mathscr]{eucal}
\usepackage{amsmath}
\usepackage{amscd}
\usepackage[dvips]{color}
\usepackage{multicol}
\usepackage[all]{xy}           %xypic macro for latex2.09
\usepackage{graphicx}
\usepackage{color}
\usepackage{colordvi}
\usepackage{xspace}
\usepackage{bbm}

\usepackage{tikz}

\begin{document}

\newcommand{\nc}{\newcommand}
\newcommand{\delete}[1]{}
\nc{\mfootnote}[1]{\footnote{#1}} % Use this to show footnotes
\nc{\todo}[1]{\tred{To do:} #1}

\delete{
\nc{\mlabel}[1]{\label{#1}}  % Use this to suppress names
\nc{\mcite}[1]{\cite{#1}}  % Use this to suppress names
\nc{\mref}[1]{\ref{#1}}  % Use this to suppress names
\nc{\mbibitem}[1]{\bibitem{#1}} % Use this to show number
}

%\delete{
\nc{\mlabel}[1]{\label{#1}  % Use the next two lines to show names
{\hfill \hspace{1cm}{\bf{{\ }\hfill(#1)}}}}
\nc{\mcite}[1]{\cite{#1}{{\bf{{\ }(#1)}}}}  % Use this lines to show names
\nc{\mref}[1]{\ref{#1}{{\bf{{\ }(#1)}}}}  % Use this lines to show names
\nc{\mbibitem}[1]{\bibitem[\bf #1]{#1}} % Use this to show name
%}

%%%%%%%%%%%%%%%%%%%%%%%% Statements
\newtheorem{thm}{Theorem}[section]
\newtheorem{lem}[thm]{Lemma}
\newtheorem{cor}[thm]{Corollary}
\newtheorem{pro}[thm]{Proposition}
\newtheorem{ex}[thm]{Example}
\newtheorem{rmk}[thm]{Remark}
\newtheorem{defi}[thm]{Definition}

\renewcommand{\labelenumi}{{\rm(\alph{enumi})}}
\renewcommand{\theenumi}{\alph{enumi}}

\nc{\tred}[1]{\textcolor{red}{#1}}
\nc{\tblue}[1]{\textcolor{blue}{#1}}
\nc{\tgreen}[1]{\textcolor{green}{#1}}
\nc{\tpurple}[1]{\textcolor{purple}{#1}}
\nc{\btred}[1]{\textcolor{red}{\bf #1}}
\nc{\btblue}[1]{\textcolor{blue}{\bf #1}}
\nc{\btgreen}[1]{\textcolor{green}{\bf #1}}
\nc{\btpurple}[1]{\textcolor{purple}{\bf #1}}

\nc{\huaA}{{\mathcal A}} \nc{\huaB}{\mathcal B}
\nc{\huaC}{{\mathcal C}} \nc{\huaD}{{\mathcal D}}
\nc{\huaE}{{\mathcal E}} \nc{\huaF}{{\mathcal F}}
\nc{\huaG}{{\mathcal G}} \nc{\huaH}{{\mathcal H}}
\nc{\huaI}{{\mathcal I}} \nc{\huaJ}{{\mathcal J}}
\nc{\huaL}{{\mathcal L}} \nc{\huaM}{{\mathcal M}}
\nc{\huaN}{{\mathcal N}} \nc{\huaO}{{\mathcal O}}
\nc{\huaP}{{\mathcal P}} \nc{\huaQ}{{\mathcal Q}}
\nc{\huaR}{{\mathcal R}} \nc{\huaS}{{\mathcal S}}
\nc{\huaT}{{\mathscr T}} \nc{\huaU}{{\mathcal U}}
\nc{\huaV}{{\mathcal V}} \nc{\huaW}{{\mathcal W}}
\nc{\huaX}{{\mathcal X}} \nc{\huaY}{{\mathcal Y}}
\nc{\huaZ}{{\mathcal Z}}
%%%%%%%%%%%%%%%%%%  frak fonts
\nc{\frka}{\mathfrak a}
\nc{\frkb}{\mathfrak b}
\nc{\frkc}{\mathfrak c}
\nc{\frkd}{\mathfrak d}
\nc{\frke}{\mathfrak e}
\nc{\frkf}{\mathfrak f}
\nc{\frkg}{\mathfrak g}
\nc{\frkh}{\mathfrak h}
\nc{\frki}{\mathfrak i}
\nc{\frkj}{\mathfrak j}
\nc{\frkk}{\mathfrak k}
\nc{\frkl}{\mathfrak l}
\nc{\frkm}{\mathfrak m}
\nc{\frkn}{\mathfrak n}
\nc{\frko}{\mathfrak o}
\nc{\frkp}{\mathfrak p}
\nc{\frkq}{\mathfrak q}
\nc{\frkr}{\mathfrak r}
\nc{\frks}{\mathfrak s}
\nc{\frkt}{\mathfrak t}
\nc{\frku}{\mathfrak u}
\nc{\frkv}{\mathfrak v}
\nc{\frkw}{\mathfrak w}
\nc{\frkx}{\mathfrak x}
\nc{\frky}{\mathfrak y}
\nc{\frkz}{\mathfrak z}

\nc{\frkA}{\mathfrak A}
\nc{\frkB}{\mathfrak B}
\nc{\frkC}{\mathfrak C}
\nc{\frkD}{\mathfrak D}
\nc{\frkE}{\mathfrak E}
\nc{\frkF}{\mathfrak F}
\nc{\frkG}{\mathfrak G}
\nc{\frkH}{\mathfrak H}
\nc{\frkI}{\mathfrak I}
\nc{\frkJ}{\mathfrak J}
\nc{\frkK}{\mathfrak K}
\nc{\frkL}{\mathfrak L}
\nc{\frkM}{\mathfrak M}
\nc{\frkN}{\mathfrak N}
\nc{\frkO}{\mathfrak O}
\nc{\frkP}{\mathfrak P}
\nc{\frkQ}{\mathfrak Q}
\nc{\frkR}{\mathfrak R}
\nc{\frkS}{\mathfrak S}
\nc{\frkT}{\mathfrak T}
\nc{\frkU}{\mathfrak U}
\nc{\frkV}{\mathfrak V}
\nc{\frkW}{\mathfrak W}
\nc{\frkX}{\mathfrak X}
\nc{\frkY}{\mathfrak Y}
\nc{\frkZ}{\mathfrak Z}

\nc{\g}{\frkg}
\nc{\Der}{\mathrm{Der}}
\nc{\Pint}{\mathbb{Z}}
\nc{\pf}{\noindent{\bf Proof.}}
\def\qed{\hfill ~\vrule height6pt width6pt depth0pt}
\nc {\emptycomment}[1]{} %to remove paragraphs
\nc{\h}{\mathbbm h}
\nc{\Fu}{\mathrm{F}}
\nc{\ad}{\mathrm{ad}}
\nc{\gl}{\mathfrak {gl}}
\nc{\Id}{\mathrm{id}}
\nc{\sgn}{\mathrm{sgn}}
\nc{\Ksgn}{\mathrm{Ksgn}}
\nc{\Real}{\mathbb R}
\nc{\Comp}{\mathbb C}
\nc{\Numb}{\mathbb N}

\font\cyr=wncyr10

\nc{\redtext}[1]{\textcolor{red}{#1}}
\nc{\cm}[1]{\textcolor{purple}{Chengming: #1}}
\nc{\li}[1]{\textcolor{red}{Li: #1}}
\nc{\jun}[1]{\textcolor{blue}{Jun: #1}}

%%%%%%%%%%%%%%%%%%%%%%%%%%%%%%%%%%%%%%%%%%%%%%%%%%%%%%%%%%%%%%%%%%

\title{Nijenhuis operators on $n$-Lie algebras}

\author{Jiefeng Liu}
\address{Department of Mathematics, Jilin University, Changchun 130012, Jilin, China}
\email{jfliu13@mails.jlu.edu.cn }

\author{Yunhe Sheng}
\address{Department of Mathematics, Jilin University, Changchun 130012, Jilin, China}
\email{shengyh@jlu.edu.cn}

\author{Yanqiu Zhou}
\address{Department of Mathematics, Jilin University, Changchun 130012, Jilin, China}
\email{yqzhou15@mails.jlu.edu.cn}

\author{Chengming Bai}
\address{Chern Institute of Mathematics \& LPMC, Nankai University, Tianjin 300071, China}
         \email{baicm@nankai.edu.cn}

%\date{\today}
%\footnotetext{{\it{Keyword}:  Nijenhuis operators, $n$-Lie algebras, deformations, Rota-Baxter operators}} \footnotetext{{\it{MSC}}: 17B99, 13D10.}

\keywords{Nijenhuis operators, $n$-Lie algebras, deformations,
Rota-Baxter operators}

\subjclass[2000]{17B99, 13D10}
\maketitle

%\begin{document}

\begin{abstract}
In this paper, we study $(n-1)$-order deformations of an $n$-Lie algebra and introduce the notion of a Nijenhuis operator on an $n$-Lie algebra, which could give rise to trivial deformations. We prove that a polynomial of a Nijenhuis operator is still a Nijenhuis operator. Finally, we give various constructions of Nijenhuis operators and some examples.
\end{abstract}

\setcounter{section}{0}
\section{Introduction}

The notion of a Filippov algebra, or an $n$-Lie algebra was introduced in  \cite{Filippov}. It is the algebraic structure corresponding to Nambu mechanics \cite{Gautheron,N,T}. $n$-Lie algebras, or more generally, $n$-Leibniz algebras, are widely studied  \cite{construction,CasasSN,CasasHigher,DT,DI,cohomology,Kasymov,NR bracket of n-Lie,Tcohomology}. In particular, 3-Lie algebras   were studied from several aspects recently \cite{baiguosheng,Realization,BaiGuo,deformation,Zhangtao}
due to applications in the Bagger-Lambert-Gustavsson theory of multiple
M2-branes \cite{BL3,BL4,BL2,HHM,P,BL1}. See the review article \cite{review} for more details.

The aim of this paper is to study $(n-1)$-order deformations of an $n$-Lie algebra. In particular, we pay special attention to a trivial deformation, which could give rise to an operator that satisfies some conditions. We call such an operator a Nijenhuis operator on an $n$-Lie algebra. It is believed that one can learn more about a mathematical object by studying its deformations \cite{Gerstenhaber}. Furthermore, Nijenhuis operators on Lie algebras play an important role in the study of integrability of nonlinear evolution equations \cite{Dorfman1993}. Deformations of $n$-Lie algebras have been studied from several aspects.
  See \cite{review,DI, deformation,Tcohomology,Zhangtao} for more details. In particular, a notion of a Nijenhuis operator on a 3-Lie algebra   was  introduced in \cite{Zhangtao} in the study of the 1-order deformations   of a 3-Lie algebra. But there are some  quite  strong conditions in this
  definition of a Nijenhuis operator.  In the case of Lie algebras, one could obtain fruitful results by considering one-parameter infinitesimal deformations, i.e. 1-order deformations. However, for $n$-Lie algebras, we believe that one should consider $(n-1)$-order deformations to obtain similar results. In \cite{deformation}, for 3-Lie algebras, the author has already considered 2-order deformations. But Nijenhuis operators   were  not studied there. Our Nijenhuis   operators are  obtained by considering an $(n-1)$-order trivial deformation of an $n$-Lie algebra. On the other hand, our Nijenhuis operators on 3-Lie algebras match  up very well with  some other existing interesting operators, such as Rota-Baxter operators \cite{BaiGuo} and $\mathcal O$-operators \cite{baiguosheng} on $3$-Lie algebras.

  The paper is organized as follows. In Section 2, we recall  some facts on  $n$-Lie algebras, their representations and associated cohomologies. In Section 3, we consider $(n-1)$-order deformations of an $n$-Lie algebra. We give the notion of a Nijenhuis operator on an $n$-Lie algebra, and show that it could give rise to a trivial deformation (Theorem \ref{thm:trivial deformation}). We show that a polynomial of a Nijenhuis operator is   still a Nijenhuis operator (Theorem \ref{th:polnij}). Furthermore, our Nijenhuis operators match up with $\mathcal O$-operators on $n$-Lie algebras (Proposition \ref{pro:ON}). In Section 4, according to constructions of $n$-Lie algebras, we give various constructions of Nijenhuis operators. We   also give examples of Nijenhuis operators on some 4-dimensional  3-Lie algebras as a guide for a further development.
  \section{Preliminaries}

\begin{defi}\label{defi of n-LA}
An $n$-Lie algebra $\g$ is a vector space together with an $n$-multilinear  skew-symmetric bracket $[\cdot,\cdots,\cdot]:\wedge^n\g\longrightarrow\g$ such that for all $x_i,y_i\in\g$, the following Filippov  identity is satisfied:
\begin{eqnarray}\label{FI}
[x_1,x_2,\cdots,x_{n-1},[y_1,y_2,\cdots,y_n]]=\sum_{i=1}^{n}[y_1,y_2,\cdots,[x_1,x_2,\cdots,x_{n-1},y_i],\cdots,y_n].
\end{eqnarray}
\end{defi}

For $x_1,x_2,\cdots,x_{n-1}\in\g$, define $\ad:\wedge^{n-1}\g\longrightarrow\gl(\g)$ by $$\ad_{x_1,x_2,\cdots,x_{n-1}}y=[x_1,x_2,\cdots,x_{n-1},y], \quad\forall y\in \g.$$
Then Eq. \eqref{FI} is equivalent to that $\ad_{x_1,x_2,\cdots,x_{n-1}}$ is a derivation, i.e.
$$
\ad_X[y_1,y_2,\cdots,y_n]=\sum_{i=1}^{n}[y_1,y_2,\cdots,\ad_Xy_i,\cdots,y_n],\quad \forall X=(x_1,x_2,\cdots,x_{n-1})\in\wedge^{n-1}\g.
$$
Elements in $\wedge^{n-1}\g$ are called {\bf fundamental objects} of the $n$-Lie algebra $(\g,[\cdot,\cdots,\cdot])$. In the sequel, we will denote $\ad_Xy$ simply by $X\circ y$.

Define a bilinear  operation on the set of fundamental objects $ \circ: (\wedge^{n-1}\g)\otimes(\wedge^{n-1}\g)\longrightarrow \wedge^{n-1}\g$  by
\begin{eqnarray}\label{comp FOS}
X\circ Y=\sum_{i=1}^{n-1}(y_1,\cdots,y_{i-1},X\circ y_i,y_{i+1},\cdots,y_{n-1}),
\end{eqnarray}
for all $X=(x_1,x_2,\cdots,x_{n-1})$ and $Y=(y_1,y_2,\cdots,y_{n-1}).$
In \cite{DT}, the authors proved that $(\wedge^{n-1}\g,\circ)$ is a Leibniz algebra, i.e. the following equality holds:
\begin{eqnarray}\label{Leibniz circle}
X\circ (Y\circ Z)=(X\circ Y)\circ Z+Y\circ (X\circ Z),\quad\forall X,Y,Z\in\wedge^{n-1}\g.
\end{eqnarray}
Moreover, the Filippov  identity (\ref{FI}) is equivalent to
\begin{eqnarray}\label{FI3}
X\circ (Y\circ z)-Y\circ (X\circ z)=(X\circ Y)\circ z,\quad\forall X,Y\in\wedge^{n-1}\g,z\in\g.
\end{eqnarray}

\begin{defi}
Let $V$ be a vector space. A representation of an $n$-Lie algebra $\frkg$ on $V$ is a multilinear map $\rho:\wedge^{n-1}\frkg\longrightarrow \gl(V),$ such that for all $\ X,Y\in\wedge^{n-1}\frkg,\ x_i,y_i\in\g$, the following equalities hold:
\begin{eqnarray*}[\rho(X),\rho(Y)]&=&\rho(X\circ Y),\\
 \rho(x_1,x_2,\cdots,x_{n-2},[y_1,y_2,\cdots,y_n])&=&\sum_{i=1}^n(-1)^{n-i}\rho(y_1,\cdots,\hat{y_i},\cdots,y_n)\rho(x_1,\cdots,x_{n-2},y_i),
 \end{eqnarray*}
 where $\hat {y_i}$ means that $y_i$ is omitted.
\end{defi}

We denote a representation by $(V;\rho)$.

Given a representation $(V;\rho)$, there is a semidirect product $n$-Lie algebra structure on $\g\oplus V$ given by
$$
[x_1+v_1,\cdots, x_n+v_n]=[x_1,\cdots,x_n]+\sum_{i=1}^n(-1)^{n-i}\rho(x_1,\cdots,\hat{x_i},\cdots,x_{n})(v_i),\quad \forall x_i\in\g,~v_i\in V.
$$

We denote this semidirect $n$-Lie algebra simply by $\g\ltimes_\rho V$. In particular, when $n=2$, i.e. for a Lie algebra, we obtain the usual notion of a semidirect product Lie algebra.

A $p$-cochain on  $\frkg$ with the coefficients in a representation   $(V;\rho)$ is  a multilinear  map $\alpha^p:\wedge^{n-1}\frkg\otimes\stackrel{(p-1)}{\ldots}\otimes\wedge^{n-1}\frkg\wedge\frkg\longrightarrow V$.
Denote the space of $p$-cochains by $C^{p}(\g;V).$ The coboundary operator $\delta:C^{p}(\g;V)\longrightarrow C^{p+1}(\g;V)$  is given by
\begin{eqnarray*}
&&(\delta\alpha^p)(X_1,\ldots,X_p,z)\\
&=& \sum_{1\leq i<k}(-1)^i\alpha^p(X_1,\cdots,\hat{X}_i,\cdots,X_{k-1},X_i\circ X_k,X_{k+1},\cdots,X_{p},z)\\
&&+\sum_{i=1}^p(-1)^i\alpha^p(X_1,\cdots,\hat{X}_i,\cdots,\frkX_{p},[X_i,z])\\
&&+\sum_{i=1}^p(-1)^{i+1}\rho(X_i)\alpha^p(X_1,\cdots,\hat{X}_i,\cdots,X_{p},z)\\
&&+\sum_{i=1}^{n-1}(-1)^{n+p-i+1}\rho(x^1_p,x^2_p,\cdots,\hat{x}^i_p,\cdots,x^{n-1}_p,z)\alpha^p(X_1,\cdots,X_{p-1},x^i_p) ,
\end{eqnarray*}
for all   $X_i=(x^1_i,x^2_i,\cdots,x^{n-1}_i)\in\wedge^{n-1}\frkg$ and $z\in\frkg.$
  \section{Nijenhuis operators on $n$-Lie algebras}

In this section, first we study $(n-1)$-order deformations of an $n$-Lie algebra, and  introduce the notion of a Nijenhuis operator on an $n$-Lie algebra, which could  generate a trivial deformation. Then we show that a polynomial of a Nijenhuis operator is still a Nijenhuis operator. Finally, we give the relation between $\mathcal O$-operators and Nijenhuis operators.

\subsection{$(n-1)$-order deformations of an $n$-Lie algebra}

Let $(\g,[\cdot,\cdots,\cdot])$ be an $n$-Lie algebra. Let $\omega_i:\otimes^n\g\longrightarrow \g,~~ 1\leq i\leq n-1$ be skew-symmetric   multilinear   maps. Consider a $\lambda$-parametrized family of linear operations:
\begin{equation}\label{eq:deformation}
[x_1,x_2,\cdots,x_{n-1},x_n]_\lambda=[x_1,x_2,\cdots,x_{n-1},x_n]+\sum_{i=1}^{n-1}\lambda^i\omega_i(x_1,x_2,\cdots,x_{n-1},x_n).
\end{equation}
Here $\lambda\in\mathbb F,$ where $\mathbb F$ is the base field. If all $[\cdot,\cdots,\cdot]_\lambda$ are  $n$-Lie algebra structures, we say that
$\omega_1,\cdots,\omega_{n-1}$ generate   {\bf an $(n-1)$-order $1$-parameter  deformation} of the $n$-Lie algebra $(\g,[\cdot,\cdots,\cdot])$.

\begin{pro}\label{conds}
With the above notations, $\omega_1,\cdots,\omega_{n-1}$ generate   an $(n-1)$-order $1$-parameter   deformation of the $n$-Lie algebra $(\g,[\cdot,\cdots,\cdot])$ if and only if the following conditions are satisfied:
  \begin{eqnarray}
   \delta\omega_1&=&0;\label{cond1}\\
   \delta\omega_l+\frac{1}{2}\sum_{i=1}^{l-1}[\omega_i,\omega_{l-i}]&=&0,\quad 2\leq l\leq n-1;\label{cond2}\\
   \frac{1}{2}\sum_{i=l-n+1}^{n-1}[\omega_i,\omega_{l-i}]&=&0,\quad n\leq l\leq 2n-2.\label{cond3}
   \end{eqnarray}
Here $[\omega_i,\omega_j]$ is given by
\begin{eqnarray}\label{N-R bracket}
[\omega_i,\omega_j](X,Y,z)&=&\omega_i(X,\omega_j(Y,z))-\omega_i(Y,\omega_j(X,z))+\omega_j(X,\omega_i(Y,z))-\omega_j(Y,\omega_i(X,z))\nonumber\\
&&-\omega_i(\omega_j(X,\cdot)\circ Y,z)-\omega_j(\omega_i(X,\cdot)\circ Y,z),\quad\forall X,Y\in\wedge^{n-1}\g,~z\in\g,
\end{eqnarray}
where $\omega_j(X,\cdot)\circ Y\in\wedge^{n-1}\g$ is given by
\begin{eqnarray*}
\omega_j(X,\cdot)\circ Y=\sum_{k=1}^{n-1}y_1\wedge\cdots\wedge\omega_j(X,y_k)\wedge\cdots\wedge y_{n-1},\quad \forall~ Y=(y_1,\cdots, y_{n-1}).
\end{eqnarray*}
\end{pro}
\pf
All $[\cdot,\cdots,\cdot]_\lambda$ are  $n$-Lie algebra structures if and only if
$$X\circ_\lambda (Y\circ_\lambda z)-(X\circ_\lambda Y)\circ_\lambda z-Y\circ_\lambda (X\circ_\lambda z)=0,\quad \forall X,~Y\in \wedge^{n-1}\g, z\in \g.$$
First we have
\begin{eqnarray*}
Y\circ_\lambda z=Y\circ z+\sum_{i=1}^{n-1}\lambda^i\omega_i(Y,z);\quad
X\circ_\lambda Y=X\circ Y+\sum_{i=1}^{n-1}\lambda^i\omega_i(X,\cdot)\circ Y.
\end{eqnarray*}
Then by direct computations, we have
\begin{eqnarray*}
X\circ_\lambda (Y\circ_\lambda z)&=&X\circ (Y\circ z)+\sum_{i=1}^{n-1}\lambda^i\big(\omega_i(X,Y\circ z)+X\circ \omega_i(Y,z)\big)\\&&+\sum_{i,j=1}^{n-1}\lambda^{i+j}\omega_i(X,\omega_j(Y,z));\\
Y\circ_\lambda (X\circ_\lambda z)&=&Y\circ (X\circ z)+\sum_{i=1}^{n-1}\lambda^i\big(\omega_i(Y,X\circ z)+Y\circ \omega_i(X,z)\big)\\&&+\sum_{i,j=1}^{n-1}\lambda^{i+j}\omega_i(Y,\omega_j(X,z));\\
(X\circ_\lambda Y)\circ_\lambda z&=&(X\circ Y)\circ z+\sum_{i=1}^{n-1}\lambda^i\big(\omega_i(X\circ Y,z)+ (\omega_i(X,\cdot)\circ Y)\circ z\big)\\&&+\sum_{i,j=1}^{n-1}\lambda^{i+j}\omega_i(\omega_j(X,\cdot)\circ Y)\circ z.
\end{eqnarray*}
Comparing the coefficients of $\lambda^l,1\leq l\leq 2n-2$, we obtain conditions (\ref{cond1})-(\ref{cond3}) respectively.\qed

\begin{rmk}
  The bracket given by Eq. $\eqref{N-R bracket}$ is just the Nijenhuis-Richardson bracket associated to   an $n$-Lie algebra. See \cite{NR bracket of n-Lie} for more details.
\end{rmk}

\begin{defi}\label{defi:trivial}
A deformation is said to be {\bf trivial} if there exists a linear map $N:\g\longrightarrow \g$ such that for all $\lambda$,  $T_\lambda={\Id}+\lambda N$ satisfies
\begin{equation}\label{eq:trivial}
  T_\lambda[x_1,x_2,\cdots,x_n]_\lambda=[ T_\lambda x_1, T_\lambda x_2, \cdots, T_\lambda x_n],\quad \forall x_1,\cdots, x_n\in \g.
\end{equation}
\end{defi}

The left hand side of Eq.~\eqref{eq:trivial} equals to
\begin{eqnarray*}
&&[x_1,x_2,\cdots,x_n]+\lambda\big(\omega_1(x_1,x_2,\cdots,x_n)+N[x_1,x_2,\cdots,x_n]\big)\\
&&+\sum_{j=1}^{n-2}\lambda^{j+1}\big(\omega_{j+1}(x_1,x_2,\cdots,x_n)+N\omega_j(x_1,x_2,\cdots,x_n)\big)+\lambda^nN\omega_{n-1}(x_1,x_2,\cdots,x_n).
\end{eqnarray*}
The right hand side of Eq.~\eqref{eq:trivial} equals to
\begin{eqnarray*}
&&[x_1,x_2,\cdots,x_n]+\lambda\sum_{i=1}^n[x_1,\cdots,Nx_i,\cdots,x_n]+\lambda^2\sum_{i<j}[x_1,\cdots,Nx_i,\cdots,Nx_j,\cdots,x_n]\\
&&+\lambda^3\sum_{i<j<k}[x_1,\cdots,Nx_i,\cdots,Nx_j,\cdots,Nx_k,\cdots,x_n]+\cdots+\lambda^n[Nx_1,Nx_2,\cdots,Nx_n].
\end{eqnarray*}
Therefore, by Eq.~\eqref{eq:trivial}, we have
\begin{eqnarray}\label{eq:trivialdef con1}
 \omega_1(x_1,x_2,\cdots,x_n)+N[x_1,x_2,\cdots,x_n]&=&\sum_{i=1}^n[x_1,\cdots,Nx_i,\cdots,x_n],\\
 \label{eq:trivialdef con2}N\omega_{n-1}(x_1,x_2,\cdots,x_n)&=&[Nx_1,Nx_2,\cdots,Nx_n],
 \end{eqnarray}
and
\begin{equation}\label{eq:trivialdef con3}
 \omega_l(x_1,x_2,\cdots,x_n)+N\omega_{l-1}(x_1,x_2,\cdots,x_n)
=\sum_{i_1<i_2\cdots<i_l}[\cdots,Nx_{i_1},\cdots,Nx_{i_k},\cdots,Nx_{i_l},\cdots],
\end{equation}
for all $2\leq l\leq n-1$.

\emptycomment{
\begin{eqnarray*}\label{eq:trivialdef con1}
 \omega_1(x_1,x_2,\cdots,x_n)+N[x_1,x_2,\cdots,x_n]&=&\sum_{i=1}^n[x_1,\cdots,Nx_i,\cdots,x_n],\\
 \omega_2(x_1,x_2,\cdots,x_n)+N\omega_1(x_1,x_2,\cdots,x_n)&=&\sum_{i<j}[x_1,\cdots,Nx_i,\cdots,Nx_j,\cdots,x_n],\\
 \omega_3(x_1,x_2,\cdots,x_n)+N\omega_2(x_1,x_2,\cdots,x_n)&=&\sum_{i<j<k}[x_1,\cdots,Nx_i,\cdots,Nx_j,\cdots,Nx_k,\cdots,x_n],\\
  \quad\quad\quad\quad\quad\quad\quad\quad\quad\quad\quad\quad\vdots  \\
 \omega_l(x_1,x_2,\cdots,x_n)+N\omega_{l-1}(x_1,x_2,\cdots,x_n)
&=&\sum_{i_1<i_2\cdots<i_l}[x_1,\cdots,Nx_{i_1},\cdots,Nx_{i_k},\cdots,Nx_{i_l},\cdots,x_n],\\
  \quad\quad\quad\quad\quad\quad\quad\quad\quad\quad\quad\quad\vdots  \\
   N\omega_{n-1}(x_1,x_2,\cdots,x_n)&=&[Nx_1,Nx_2,\cdots,Nx_n].
\end{eqnarray*}
}

Let $(\g,[\cdot,\cdots,\cdot])$ be an $n$-Lie algebra, and $N:\g\longrightarrow\g$ a linear map. Define an $n$-ary bracket $[\cdot,\cdots,\cdot]_N^1:\wedge^n\g\longrightarrow\g$  by
   \begin{equation}\label{eq:bracket(1)}
    [x_1,x_2,\cdots,x_n]_N^{1}=\sum_{i=1}^n[x_1,\cdots,Nx_i,\cdots,x_n]-N[x_1,x_2,\cdots,x_n].
  \end{equation}
  Then we define $n$-ary brackets $[\cdot,\cdots,\cdot]_N^j:\wedge^n\g\longrightarrow\g, (2\leq j\leq n-1)$ via induction by
    \begin{equation}\label{eq:bracket (j)}
    [x_1,x_2,\cdots,x_n]_N^{j}=\sum_{i_1<i_2\cdots<i_{j}}[\cdots,Nx_{i_1},\cdots,Nx_{i_{j}},\cdots]-N[x_1,x_2,\cdots,x_n]_N^{j-1}.
  \end{equation}
  In particular, we have
    \begin{equation}\label{eq:bracket(n-1)}
    [x_1,x_2,\cdots,x_n]_N^{n-1}=\sum_{i_1<i_2\cdots<i_{n-1}}[\cdots,Nx_{i_1},\cdots,Nx_{i_{n-1}},\cdots]-N[x_1,x_2,\cdots,x_n]_N^{n-2}.
  \end{equation}

\begin{defi}\label{defi:Nijenhuis}
Let $(\g,[\cdot,\cdots,\cdot])$ be an $n$-Lie algebra. A linear map    $N:\g\longrightarrow\g$   is called a {\bf Nijenhuis operator} if
\begin{equation}\label{eq:Nijenhuis(n)}
[Nx_1,Nx_2,\cdots,Nx_n]=N( [x_1,x_2,\cdots,x_n]_N^{n-1}),\quad\forall x_1,\cdots, x_n\in \g.
\end{equation}
\end{defi}

Note that when $n=2$, i.e. for a Lie algebra, we obtain the usual notion of a Nijenhuis operator on a Lie algebra. More precisely, a linear transformation $N:\g\longrightarrow\g$ is a Nijenhuis operator on the Lie algebra $(\g,[\cdot,\cdot])$ if the following equality holds:
$$
[Nx,Ny]=N([Nx,y]+[x,Ny]-N[x,y]),\quad\forall x,y\in\g.
$$

\begin{rmk}
In \cite{review,DI, cohomology}, the authors considered deformations of the form  $[\cdot,\cdots,\cdot]+\lambda\omega(\cdot,\cdots,\cdot)$. In \cite{deformation}, for $3$-Lie algebras, the author has considered deformations of  the form of Eq.~\eqref{eq:deformation}. But Nijenhuis operators  were not considered there.
 On the other hand,
 in \cite{Zhangtao}, the author has introduced a notion of a Nijenhuis operator on a $3$-Lie algebra in the study of $1$-order trivial deformations.
In that definition, there is a quite strong condition $[Nx_1,Nx_2,N_3]=0$, whereas the above definition for $n=3$ is $[Nx_1,Nx_2,N_3]=N( [x_1,x_2,x_3]_N^{2})$.
  So obviously, the above definition is different with
these studies.

\emptycomment{Note that in the case of Lie algebras, we only need to consider $[\cdot,\cdot]+\lambda\omega(\cdot,\cdot)$ for a  $1$-parameter infinitesimal deformation. However, in the case of $3$-Lie algebras, it is not enough to only consider $[\cdot,\cdot]+\lambda\omega(\cdot,\cdot)$. We believe that Eq.~\eqref{eq:deformation} is the right form of  $1$-parameter infinitesimal deformation of $3$-Lie algebras, which will produce many interesting structures, see Theorem \ref{thm:trivial deformation} and Proposition \ref{pro:ON}.}
\end{rmk}

By Eqs.~\eqref{eq:bracket (j)} and \eqref{eq:Nijenhuis(n)}, we have
\begin{pro}
Let $(\g,[\cdot,\cdots,\cdot])$ be an $n$-Lie algebra.  Then $N$ is a Nijenhuis operator on $\g$ if and only if $N$ satisfies the following equality
\begin{equation}\label{eq:Nijenhuis all}
 \sum_{p=0}^n\sum_{\sigma}(-1)^{\frac{p(p-1)}{2}+\sum_{j=1}^p\sigma(j)}N^p[x_{\sigma(1)},\cdots,x_{\sigma(p)},Nx_{\sigma(p+1)},\cdots,Nx_{\sigma(n)}]=0,\quad\forall x_i\in\g,
 \end{equation}
 where the summation is taken over all $(p,n-p)$-unshuffles, i.e. $\sigma(1)<\cdots<\sigma(p),~\sigma(p+1)<\cdots<\sigma(n)$.
\end{pro}

By Eqs.~\eqref{eq:trivialdef con1}-\eqref{eq:trivialdef con3}, it is straightforward to deduce that a trivial deformation gives rise to a Nijenhuis operator. The following theorem shows that the converse is also true.

\begin{thm}\label{thm:trivial deformation}
  Let $N$ be a Nijenhuis operator on an $n$-Lie algebra
$(\g,[\cdot,\cdots,\cdot])$. Then
   a deformation can be obtained by putting
  \begin{equation}\label{eq:formulaOT}
    \omega_i(x_1,x_2,\cdots,x_n)=[x_1,x_2,\cdots,x_n]_N^{i},\quad 1\leq i\leq n-1.
  \end{equation}
  Moreover, this deformation  is trivial.
\end{thm}

One way to prove  this theorem directly is to verify that conditions in Proposition \ref{conds} are satisfied. Instead we apply a different method to prove this theorem to avoid complicated and lengthy computations. The following general fact is an important ingredient in the proof.

\begin{lem}\label{lem:isomorphism}
  Let $(\g,[\cdot,\cdots,\cdot])$ be an $n$-Lie algebra and $\h$ a vector space with an $n$-ary bracket $[\cdot,\cdots,\cdot]'$. If there exists an isomorphism between vector spaces, say $f:\h\longrightarrow \g$, such that
  $$
  f[x_1,x_2,\cdots,x_n]'=[f(x_1),f(x_2),\cdots,f(x_n)],\quad\forall ~x_i\in\h,
$$
then $(\h,[\cdot,\cdots,\cdot]')$ is an $n$-Lie algebra.
\end{lem}
\pf It follows from
straightforward computations.   \qed\vspace{3mm}

%\pf Since $(\g,[\cdot,\cdot,\cdot])$ is a $3$-Lie algebra, we have
%\begin{eqnarray*}
%0&=&[f(x_1),f(x_2),[f(x_3),f(x_4),f(x_5)]]-[[f(x_1),f(x_2),f(x_3)],f(x_4),f(x_5)]\\
%&&-[f(x_3),[f(x_1),f(x_2),f(x_4)],f(x_5)]-[f(x_3),f(x_4),[f(x_1),f(x_2),f(x_5)]]\\
%&=&f\Big([x_1,x_2,[x_3,x_4,x_5]]-[[x_1,x_2,x_3],x_4,x_5]\\
%&&-[x_3,[x_1,x_2,x_4],x_5]-[x_3,x_4,[x_1,x_2,x_5]]\Big).
%\end{eqnarray*}
%Since $f$ is an isomorphism between vector spaces, we have
%$$[x_1,x_2,[x_3,x_4,x_5]]-[[x_1,x_2,x_3],x_4,x_5]-[x_3,[x_1,x_2,x_4],x_5]-[x_3,x_4,[x_1,x_2,x_5]]=0,$$
%which implies that $(\h,[\cdot,\cdot,\cdot]')$ is a 3-Lie algebra.
%\qed\vspace{3mm}

{\bf The proof of Theorem \ref{thm:trivial deformation}:}
It is obvious that for a Nijenhuis   operator $N$,
the maps $\omega_1,\cdots, \omega_{n-1}$ given by Eq.~\eqref{eq:formulaOT} satisfy Eq.~\eqref{eq:Nijenhuis(n)}. Therefore,
for any $\lambda$, $T_\lambda$ satisfies $$T_\lambda[x_1,x_2,\cdots,x_n]_\lambda=[T(x_1),T(x_2),\cdots,T(x_n)],\quad\forall x_1,\cdots, x_n\in \g.$$
 For $\lambda$ sufficiently small, we see that $T_\lambda$ is an
isomorphism between vector spaces. By Lemma \ref{lem:isomorphism},
we deduce that $(\g,[\cdot,\cdots,\cdot]_\lambda)$ is an $n$-Lie
algebra, for $\lambda$ sufficiently small. Thus, $\omega_1,\cdots, \omega_{n-1}$
  given by
Eq.~\eqref{eq:formulaOT} satisfy the conditions
(\ref{cond1})-(\ref{cond3}) in Proposition \ref{conds}. Therefore,
$(\g,[\cdot,\cdots,\cdot]_\lambda)$ is an $n$-Lie algebra for all
$\lambda$, which means that $\omega_1,\cdots, \omega_{n-1}$
given by
Eq.~\eqref{eq:formulaOT} generate a deformation. It is obvious  that
this deformation is trivial.\qed

\begin{cor}\label{cor:n-LA}
  Let $N$ be a Nijenhuis operator on an $n$-Lie algebra
$(\g,[\cdot,\cdots,\cdot])$.
  Then $(\g,[\cdot,\cdots,\cdot]_N^{n-1})$ is an $n$-Lie algebra, and $N$ is a   homomorphism from $(\g,[\cdot,\cdots,\cdot]_N^{n-1})$ to $(\g,[\cdot,\cdots,\cdot])$.
\end{cor}

At the end of this subsection,   as an example, we study Nijenhuis operators on 3-dimensional  complex 3-Lie algebras. Recall that a linear map $N$ acting on a $3$-Lie algebra $(\g,[\cdot,\cdot,\cdot])$ is a  Nijenhuis operator if
\begin{equation}\label{eq:3Nijenhuis}
[Nx,Ny,Nz]=N( [x,y,z]_N^{2}),
\end{equation}
where the $3$-ary bracket $[\cdot,\cdot,\cdot]_N^{2}$ is defined by
  \begin{equation}\label{eq:bracketNN}
    [x,y,z]_N^{2}=[N x,Ny,z]+[ x,Ny,Nz]+[ Nx,y,Nz]-N[ x,y,z]_N^1,
  \end{equation}
where the $3$-ary bracket $[\cdot,\cdot,\cdot]_N^1$ is defined by
  \begin{equation}\label{eq:bracketN}
    [x,y,z]_N^1=[N x,y,z]+[ x,Ny,z]+[ x,y,Nz]-N[ x,y,z].
  \end{equation}

 It is obvious that any linear transformation on an abelian 3-Lie algebra is a Nijenhuis operator.
On the other hand, it is known that up to isomorphism, there is only one 3-dimensional non-abelian   complex $3$-Lie
algebra      given by
\begin{equation}\label{eq:3-dim}
[e_1,e_2,e_3]=e_1,
\end{equation}
where  $\{e_1 ,e_2 ,e_3\}$ is a basis of $\g$.

\begin{thm}
Let $(\g,[\cdot,\cdot,\cdot])$ be the $3$-dimensional   complex $3$-Lie algebra given above. Then any
  linear transformation $N$ on $\g$
 is a Nijenhuis operator.
\end{thm}
\pf Assume
$
Ne_i=N_i^je_j.
$
  Then we have
\begin{eqnarray*}
[e_1,e_2,e_3]^1_N&=&N_1^1e_1+N_2^2e_1+N_3^3e_1-N_1^je_j,\\
N[e_1,e_2,e_3]^1_N&=&N_1^1N_1^je_j+N_2^2N_1^je_j+N_3^3N_1^je_j-N_1^jN_j^ke_k.
\end{eqnarray*}
Furthermore, we have
\begin{eqnarray*}
[e_1,e_2,e_3]^2_N&=&(N_2^2N_3^3-N_2^3N_3^2)e_1+(N_1^3N_3^2-N_3^3N_1^2)e_2+(N_1^2N_2^3-N_2^2N_1^3)e_3,\\
N[e_1,e_2,e_3]^2_N&=&\big( N_1^1(N_2^2N_3^3-N_2^3N_3^2)+N_2^1(N_1^3N_3^2-N_3^3N_1^2)+N_3^1(N_1^2N_2^3-N_2^2N_1^3)\big)e_1\\
&&+\big( N_1^2(N_2^2N_3^3-N_2^3N_3^2)+N_2^2(N_1^3N_3^2-N_3^3N_1^2)+N_3^2(N_1^2N_2^3-N_2^2N_1^3) \big) e_2\\
&&+\big( N_1^3(N_2^2N_3^3-N_2^3N_3^2)+N_2^3(N_1^3N_3^2-N_3^3N_1^2)+N_3^3(N_1^2N_2^3-N_2^2N_1^3) \big) e_3\\
&=&\big( N_1^1(N_2^2N_3^3-N_2^3N_3^2)+N_2^1(N_1^3N_3^2-N_3^3N_1^2)+N_3^1(N_1^2N_2^3-N_2^2N_1^3)\big)e_1.
\end{eqnarray*}
However, we have
\begin{eqnarray*}
&&[Ne_1,Ne_2,Ne_3]\\
&=&\big( N_1^1N_2^2N_3^3-N_1^1N_2^3N_3^2+N_2^1N_1^3N_3^2-N_2^1N_3^3N_1^2+N_3^1N_1^2N_2^3-N_3^1N_2^2N_1^3\big)e_1.
\end{eqnarray*}
Therefore, we have
$$[Ne_1,Ne_2,Ne_3]=N[e_1,e_2,e_3]^2_N.$$
The proof is finished.\qed
\subsection{Some properties of Nijenhuis operators}

\begin{lem}\label{lem:impotant relation}
Let $N:\g\longrightarrow\g$ be a Nijenhuis operator
on an $n$-Lie algebra $\g$. For all $x_1,x_2,\cdots,x_n\in\g$ and arbitrary positive number $\alpha_1,\alpha_2,\cdots,\alpha_n\in\Pint$ there holds
\begin{eqnarray}\label{eq:impotant relation}
 &&\sum_{p=0}^n\sum_{\sigma}(-1)^{\frac{p(p-1)}{2}+\sum_{j=1}^p\sigma(j)}N^{\sum_{j=1}^p\alpha_{\sigma(j)}}\nonumber\\&&[x_{\sigma(1)},x_{\sigma(2)},\cdots,x_{\sigma(p)},
 N^{\alpha_{\sigma(p+1)}}x_{\sigma(p+1)},\cdots,N^{\alpha_{\sigma(n)}}x_{\sigma(n)}]=0,
 \end{eqnarray}
 where the summation is taken over all $(p,n-p)$-unshuffles. If $N$ is invertible, this formula is valid for arbitrary $\alpha_1,\alpha_2,\cdots,\alpha_n\in\Pint$.
\end{lem}
\pf The proof is lengthy and nontrivial.  We put it in Appendix for self-contained.\qed

\begin{thm}\label{th:polnij}
Let $N:\g\longrightarrow \g$ be a Nijenhuis operator on an $n$-Lie algebra $\g$. Then for any polynomial $P(z)=\sum^n_{i=0}c_iz^i$, the operator $P(N)$ is also a Nijenhuis operator. Furthermore, if $N$ is invertible, for any $Q(z)=\sum^n_{i=-m}c_iz^i$, $Q(N)$ is also a Nijenhuis operator.
\end{thm}
\pf For all $x_1,x_2,\cdots,x_n\in\g$, by Eq. \eqref{eq:impotant relation}, we have
\begin{eqnarray*}
&&\sum_{p=0}^n\sum_{\sigma}(-1)^{\frac{p(p-1)}{2}+\sum_{j=1}^p\sigma(j)}P(N)^p[x_{\sigma(1)},x_{\sigma(2)},\cdots,x_{\sigma(p)},P(N)x_{\sigma(p+1)},\cdots,P(N)x_{\sigma(n)}]\\
&=&\sum_{\alpha_i,1\leq i\leq n}\prod_{1\leq i\leq n} c_{\alpha_i}\Big(\sum_{p=0}^n\sum_{\sigma}(-1)^{\frac{p(p-1)}{2}+\sum_{j=1}^p\sigma(j)}N^{\sum_{j=1}^p\alpha_{\sigma(j)}}\\&&[x_{\sigma(1)},x_{\sigma(2)},\cdots,x_{\sigma(p)},
 N^{\alpha_{\sigma(p+1)}}x_{\sigma(p+1)},\cdots,N^{\alpha_{\sigma(n)}}x_{\sigma(n)}]\Big)\\
 &=&0.
\end{eqnarray*}
Therefore, $P(N)$ is a Nijenhuis operator. The second statement can be proved similarly.\qed\vspace{3mm}

\begin{rmk}
In some sense, the above property ``characterize" a Nijenhuis operator, whereas some known operators like derivations and Rota-Baxter operators
do not have such a property.
\end{rmk}

In the sequel, we give the relation between $\mathcal O$-operators and Nijenhuis operators.

\begin{defi}
 Let $(\g,[\cdot,\cdots,\cdot])$ be an $n$-Lie algebra and $(V;\rho)$ a representation. A linear map $T:V\rightarrow \g$ is called an {\bf $\mathcal O$-operator} if for all $v_1,v_2,\ldots,v_n\in V$,
\begin{eqnarray}\label{eq:O-operator}
[Tv_1,Tv_2,\cdots,Tv_n]=\sum_{i=1}^n(-1)^{n-i}T\big(\rho(Tv_1,\cdots,\widehat{Tv_i},\cdots,Tv_{n})(v_i)\big).
\end{eqnarray}

\end{defi}
In particular, if we take the adjoint representation, then an $\mathcal O$-operator  is exactly a Rota-Baxter operator of weight $0$ given in \cite{BaiGuo}.

In the case of Lie algebras, the notion of an $\mathcal O$-operator was introduced by Kupershmidt in \cite{Kupershmidt} in the study of classical Yang-Baxter equation. It is straightforward to deduce that given a representation $\rho:\g\longrightarrow\gl(V)$, an $\mathcal O$-operator $T:V\longrightarrow \g$  on a Lie algebra $\g$ could give rise to a Nijenhuis operator  $\left(\begin{array}{cc}0&T\\
  0&0\end{array}\right)$  on the semidirect product Lie algebra $\g\ltimes_\rho V$.

Similarly, we have
\begin{pro}\label{pro:ON}
Let $(\g,[\cdot,\cdots,\cdot])$ be an $n$-Lie algebra and $(V;\rho)$ a representation.  A linear operator $T:V\rightarrow \g$ is   an {\bf $\mathcal O$-operator} if and only if $$\overline{T}=\left(\begin{array}{cc}0&T\\
  0&0\end{array}\right):\g\oplus V\longrightarrow \g\oplus V$$  is a Nijenhuis operator acting on the semidirect product $n$-Lie algebra $\g\ltimes_\rho V.$
\end{pro}
\pf For all $x_i\in \g,~ v_i\in V,~i=1,2,\ldots,n$, we have
$$
[\overline{T}(x_1+v_1),\cdots,\overline{T}(x_n+v_n)]=[Tv_1,\cdots,Tv_n].
$$
On the other hand, since $\overline{T}^2=0$, we have
\begin{eqnarray*}
&&\overline{T}[x_1+v_1,\cdots,x_n+v_n]_{\overline{T}}^{n-1}\\
 &=&\overline{T}\big(\sum_{i_1<i_2\cdots<i_{n-1}}[\cdots,\overline{T}(x_{i_1}+v_{i_1}),\cdots,\overline{T}(x_{i_{n-1}}+v_{i_{n-1}}),\cdots]\big)\\
 &&-\overline{T}^2 [x_1+v_1,\cdots,x_n+v_n]_{\overline{T}}^{n-2}\\
 &=& \overline{T}\big([Tv_1,Tv_2,\cdots,Tv_{n-1},x_n]+c.p.+[Tv_1,Tv_2,\cdots,Tv_{n-1},v_n]+c.p.\big)\\
 &=&T\big(\sum_{i=1}^n(-1)^{n-i}\rho(Tv_1,\cdots,\widehat{Tv_i},\cdots,Tv_{n})(v_i)\big),
\end{eqnarray*}
which implies  that  $\overline{T}$ is a Nijenhuis operator if and only if
Eq.~\eqref{eq:O-operator} is satisfied.
 \qed

\begin{rmk} In fact, when $n=2$, it is exactly the formerly mentioned conclusion for Lie algebras.
Thus,
from this point of view, our Nijenhuis operator on an $n$-Lie algebra is a natural generalization of the one on a Lie algebra,
whereas the other so-called Nijenhuis operators (for example, the ones in \cite{Zhangtao}) do not have this property.
\end{rmk}

 \section{Constructions of  Nijenhuis operators}

 \subsection{Constructions of  Nijenhuis operators on $(n+1)$-Lie algebras from  those on  $n$-Lie algebras}

 In \cite{construction}, the authors  constructed an $(n+1)$-Lie algebra $\g_f$ from an $n$-Lie algebras $\g$ using a linear function $f$. In this subsection, we show that a Nijenhuis operator on $\g$ is also a Nijenhuis operator on the  $(n+1)$-Lie algebra $\g_f$.
\begin{lem}\rm{\cite{construction}}\label{lem:const nL from n-1L}
Let $(\g,[\cdot,\cdots,\cdot])$ be an $n$-Lie algebra and $\g^*$  the dual space of $\g$. Suppose $f\in\g^*$ satisfying $f([x_1,\cdots,x_n])=0$ for all $x_i\in\g$. Then there is an $(n+1)$-Lie algebra structure on $\g$ given by
\begin{equation}
\{x_1,\cdots,x_{n+1}\}=\sum_{i=1}^{n+1}(-1)^{i-1}f(x_i)[x_1,\cdots,\hat{x_i},\cdots,x_n],\quad \forall x_i\in\g.
\end{equation}
We denote it by $\g_f$.
\end{lem}
\begin{thm}\label{pro:NijtoNij}
Assume that $N$ is a Nijenhuis operator on an $n$-Lie algebra $(\g,[\cdot,\cdots,\cdot])$. Then $N$ is also a Nijenhuis operator on the $(n+1)$-Lie algebra $(\g_f,\{\cdot,\cdots,\cdot\})$.
\end{thm}
\pf First, for $1\leq i\leq n-1$, we have
\begin{eqnarray}
\nonumber\{x_1,x_2,\cdots,x_{n+1}\}_{N}^i
&=&\sum_{j=1}^{n+1}(-1)^{i-1}\big(f(N x_j){[x_1,x_2,\cdots,\hat{Nx_j},\cdots,x_{n+1}]_N^{i-1}}\\
&&+f(x_j)[x_1,x_2,\cdots,\hat{x}_j,\cdots,x_{n+1}]_N^i\big).\label{eq:bracketrel}
 \end{eqnarray}
This fact can be proved by induction on $i$. For $i=1$, we have
\begin{eqnarray*}
&&\{x_1,x_2,\cdots,x_{n+1}\}_{N}^1\\
&=&\sum_{i=1}^{n+1}\{x_1,\cdots,Nx_i,\cdots,x_{n+1}\}-N\{x_1,x_2,\cdots,x_{n+1}\}\\
&=&\sum_{i,j,i\neq j}(-1)^{j-1}f(x_j)[x_1,\cdots,\hat{x_j},\cdots,Nx_i,\cdots,x_{n+1}]\\
&&+\sum_{i=1}^{n+1}(-1)^{i-1}f(Nx_i)[x_1,x_2,\cdots,\hat{Nx_i},\cdots,x_{n+1}]\\
&&-\sum_{i=1}^{n+1}(-1)^{i-1}f(x_i)[x_1,x_2,\cdots,\hat{x_i},\cdots,x_{n+1}]\\
&=&\sum_{i}(-1)^{i-1}\big(f(Nx_i){[x_1,x_2,\cdots,\hat{Nx_i},\cdots,x_{n+1}]}+f(x_i)[x_1,x_2,\cdots,\hat{x_i},\cdots,x_{n+1}]_N^1\big).
 \end{eqnarray*}
Now we assume that Eq.~\eqref{eq:bracketrel} holds for arbitrary $i$. Then for $i+1$, we have
\begin{eqnarray*}
 &&\{x_1,x_2,\cdots,x_{n+1}\}_N^{i+1}\\
 &=&\sum_{j_1<j_2\cdots<j_{i+1}}\{\cdots,Nx_{j_1},\cdots,Nx_{j_k},\cdots,Nx_{j_{i+1}},\cdots\}-N\{x_1,x_2,\cdots,x_n\}_{N}^i\\
 &=&\sum_{j_1<j_2\cdots<j_{i+1}}\sum_{k\neq j_1,\cdots,j_{i+1}}(-1)^{k-1}f(x_{k})[\cdots,Nx_{j_1},\cdots,\hat{x}_k,\cdots,Nx_{j_k},\cdots,Nx_{j_{i+1}},\cdots]\\
 &&+\sum_{j_1<j_2\cdots<j_{i+1}}\sum_{k= j_1,\cdots,j_{i+1}}(-1)^{k-1}f(Nx_{k})[\cdots,Nx_{j_1},\cdots,,Nx_{k},\cdots,Nx_{j_{i+1}},\cdots]\\
 &&-\sum_{k}(-1)^{k-1}N\big(f(N x_k){[x_1,x_2,\cdots,\hat{Nx_k},\cdots,x_{n+1}]_N^{i-1}}+f(x_k)[x_1,x_2,\cdots,\hat{x_k},\cdots,x_{n+1}]_N^i\big)\\
&=&\sum_{j=1}^{n+1}(-1)^{j-1}\big(f(N x_j){[x_1,x_2,\cdots,\hat{Nx_j},\cdots,x_{n+1}]_N^{i}}+f(x_j)[x_1,x_2,\cdots,\hat{x}_j,\cdots,x_{n+1}]_N^{i+1}\big),
 \end{eqnarray*}
which implies that Eq.~\eqref{eq:bracketrel} holds.
In particular, we have
\begin{eqnarray*}
 &&{\{x_1,x_2,\cdots,x_{n+1}\}}_N^{n-1}\\
&=&\sum_{j=1}^{n+1}(-1)^{j-1}\big(f(N x_j){[x_1,x_2,\cdots,\hat{Nx_j},\cdots,x_{n+1}]_N^{n-2}}+f(x_j)[x_1,x_2,\cdots,\hat{x_j},\cdots,x_{n}]_N^{n-1}\big).
 \end{eqnarray*}
 Since $N$ is a Nijenhuis operator on the $n$-Lie algebra $\g$, we have
\begin{eqnarray*}
&&\{x_1,x_2,\cdots,x_{n+1}\}_N^{n}\\
 &=&\sum_{j_1<j_2\cdots<j_{n}}\{\cdots,Nx_{j_1},\cdots,Nx_{j_k},\cdots,Nx_{j_{n}},\cdots\}-N\{x_1,x_2,\cdots,x_{n+1}\}_N^{n-1}\\
 &=&\sum_{j_1<j_2\cdots<j_{n}}\{\cdots,Nx_{j_1},\cdots,Nx_{j_k},\cdots,Nx_{j_{n}},\cdots\}\\
 &&-N\sum_{i=1}^{n+1}(-1)^{i-1}\big(f(N x_i){[x_1,x_2,\cdots,\hat{Nx_i},\cdots,x_{n+1}]_N^{n-2}}+f(x_i)[x_1,x_2,\cdots,\hat{x_i},\cdots,x_{n}]_N^{n-1}\big)\\
 &=&\sum_{i=1}^{n+1}(-1)^{i-1}f(N x_i){[x_1,x_2,\cdots,\hat{Nx_i},\cdots,x_{n+1}]_N^{n-1}}.
 \end{eqnarray*}
Furthermore, we can get \begin{eqnarray*}
N\{x_1,x_2,\cdots,x_{n+1}\}_N^{n}
&=&\sum_{i=1}^{n+1}(-1)^{i-1}f(N x_i)N{[x_1,x_2,\cdots,\hat{Nx_i},\cdots,x_{n+1}]_N^{n-1}}\\
&=&\sum_{i=1}^{n+1}(-1)^{i-1}f(N x_i)N{[Nx_1,Nx_2,\cdots,\hat{Nx_i},\cdots,Nx_{n+1}]_N^{n}}\\
&=&\{Nx_1,Nx_2,\cdots,Nx_{n+1}\},
 \end{eqnarray*}
which implies that $N$ is   a Nijenhuis operator on the $(n+1)$-Lie algebra $(\g_f,\{\cdot,\cdots,\cdot\})$.\qed

%In particular, given a Lie algebra $(\g,[\cdot,\cdot])$ and an $f\in\g^*$ satisfying $f([x,y])=0$ for all $x, y\in\g$, there is a $3$-Lie algebra structure on %$\g$ given by
%\begin{equation}\label{eq:bracket 3LA}
%\{x,y,z\}=f(x)[y,z]-f(y)[x,z]+f(z)[x,y],\forall x,y,z\in\g.
%\end{equation}
%Let $P$ be a polynomial, then for any Nijenhuis operator $N$ on $(\g,[\cdot,\cdot])$, $P(N)$ is also a Nijenhuis operator. Therefore, we have
%\begin{cor}
%Let $N$ be a Nijenhuis operator on the Lie algebra $(\g,[\cdot,\cdot])$. Then $P(N)$ is a Nijenhuis operators on the $3$-Lie algebra %$(\g,\{\cdot,\cdot,\cdot\})$.
%\end{cor}
\subsection{Constructions of  Nijenhuis operators  on $3$-Lie algebras from Nijenhuis operators on commutative associative algebras}
\emptycomment{Let $\g$ be a vector space with a bilinear product $\ast$ satisfying
\begin{equation}
(x\ast y)\ast z-x\ast(y\ast z)=(y\ast x)\ast z-y\ast(x\ast x),\quad \forall x,y,z\in\g.
\end{equation}
Then $(\g,\ast)$ is called a {\bf left-symmetric algebra}. It is obvious that all associative algebras are left-symmetric algebras. For a left-symmetric algebra $(\g,\ast)$, the commutator
\begin{equation}
[x,y]=x\ast y-y\ast x,
\end{equation}
defines a Lie algebra $ \g^c=(\g,[\cdot,\cdot])$, called the {\bf subjacent Lie algebra} of the left-symmetric algebra $(\g,\ast)$.

Let $(\g,\cdot)$ be an associative algebra, and $\omega:\g\otimes\g\rightarrow\g$ a bilinear operation. Consider a $t$-parameterized family
of multiplications $\cdot_t:\g\otimes\g\rightarrow \g$ given by
\begin{eqnarray*}
x\cdot_t y=x\cdot y+t\omega(x,y),\quad \forall x,y\in\g.
\end{eqnarray*}
If  $\g_t=(\g,\cdot_t)$ is an associative algebra for all $t\in \Real$, we say that $\omega$ generates a {\bf $1$-parameter infinitesimal
deformation} of $(\g,\cdot)$. It is easy to see that $(\g,\cdot_t)$ is
a deformation of $(\g,\cdot)$  if and only if
  \begin{eqnarray}
x\cdot\omega(y,z)+\omega(x,y\cdot z)=\omega(x,y)\cdot z+\omega(x\cdot y,z),
\end{eqnarray}
and
\begin{eqnarray}
\omega(\omega(x,y),z)=\omega(x,\omega(y,z)).
\end{eqnarray}

A deformation is said to be {\bf trivial} if there exists a linear map $N:\g\longrightarrow \g$ such that for all $t\in\mathbb R$, ${\Id}+t N$ satisfies
\begin{equation}
 ({\Id}+t N)(x\cdot_t y)= ({\Id}+t N)x\cdot ({\Id}+t N)y.
\end{equation}

As we have
\begin{eqnarray*}
({\Id}+t N)(x\cdot_t y)=x\cdot y+t(\omega(x,y)+N(x\cdot y))+t^2N\omega(x,y),
\end{eqnarray*}
and
\begin{eqnarray*}
({\Id}+t N)x\cdot ({\Id}+t N)y=x\cdot y+t(x\cdot Ny+Nx\cdot y)+t^2Nx\cdot Ny.
\end{eqnarray*}
Thus, the triviality of a deformation is equivalent to the conditions:
\begin{eqnarray}
\omega(x,y)&=&x\cdot Ny+Nx\cdot y-N(x\cdot y),\\
N\omega(x,y)&=&Nx\cdot Ny.
\end{eqnarray}

A linear map $N$ acting on an associative algebra $(\g,\cdot)$ is called a {\bf Nijenhuis operator} if
\begin{equation}\label{eq:Nijenhuis LA}
Nx\cdot Ny=N(Nx\cdot y+x\cdot Ny-N (x\cdot y)),\quad \forall~x,y\in\g.
\end{equation}

\begin{pro}
Let $N:\g\longrightarrow\g$ be a Nijenhuis operator. Then a deformation of $\g$ can be obtained by putting
$$\omega(x,y)=x\cdot Ny+Nx\cdot y-N(x\cdot y).$$
Furthermore, this deformation is a trivial one.
\end{pro}

Define
\begin{equation}
x\ast y=x\cdot Ny+Nx\cdot y-N(x\cdot y),\quad x,y\in\g.
\end{equation}

\begin{cor}
Let $N:\g\longrightarrow\g$ be a Nijenhuis operator. Then $(\g,\ast)$ is also an associative algebra, and $N$ is a morphism from $(\g,\ast)$ to $(\g,\cdot)$.
\end{cor}
}

 In fact, there is a similar study on the Nijenhuis operators on associative algebras. Explicitly,
a linear map $N$ acting on an associative algebra $(\g,\cdot)$ is called a {\bf Nijenhuis operator} if
\begin{equation}\label{eq:Nijenhuis LA}
Nx\cdot Ny=N(Nx\cdot y+x\cdot Ny-N (x\cdot y)),\quad \forall~x,y\in\g.
\end{equation}

\emptycomment{It is easy to see that if $N$ is a Nijenhuis operator on $(\g,\ast)$, $N$ is also a Nijenhuis operator on the sub-adjacent Lie algebra $\g^c$.

\begin{rmk}
More relations between Nijenhuis operators on left-symmetric algebras and   Nijenhuis operators on Lie algebras will be discussed in our future work.
\end{rmk}}

\begin{lem}\label{lem:comm asso alg}{\rm(\cite{BaiGuo})}
Let $(\g,\cdot)$ be a commutative associative algebra. Let $D\in\Der(\g)$ and $f\in \g^*$ satisfy
$f(D(x)\cdot y)=f(x\cdot D(y)).$
Then $(\g,\llbracket \cdot,\cdot,\cdot\rrbracket)$ is a $3$-Lie algebra, where the bracket is given by
\begin{eqnarray}\label{eq:Rota-Baxter 1}
\llbracket x,y,z\rrbracket&\triangleq& \begin{vmatrix}f(x)& f(y)&f(z)\\ D(x)&D(y)&D(z)\\ x&y&z\end{vmatrix}
\triangleq f(x)(D(y)\cdot z-D(z)\cdot y)+c.p..
\end{eqnarray}
\end{lem}

\begin{pro}
With the same assumptions as  Lemma \ref{lem:comm asso alg}. Let  $N$ be a Nijenhuis operator on $(\g,\cdot)$ satisfying $DN=ND$. Then  $N$ is a Nijenhuis operator on the $3$-Lie algebra $(\g,\llbracket \cdot,\cdot,\cdot\rrbracket)$, where the bracket is given by Eq. \eqref{eq:Rota-Baxter 1}.
\end{pro}
\pf  For all $x,y\in\g$, define
$[x,y]_D=D(x)\cdot y-D(y)\cdot x.$
By direct calculations,   we can verify that $(\g,[\cdot,\cdot]_D)$ is a Lie algebra. Furthermore, assume that $N$ is a Nijenhuis operator on $(\g,\cdot)$ satisfying $DN=ND$. Then  we have
\begin{eqnarray*}
[Nx,Ny]_D&=&DNx\cdot Ny-Nx\cdot DNy\\
&=&NDx\cdot Ny-Nx\cdot NDy\\
&=&N(Dx\cdot Ny+NDx\cdot y-N(Dx\cdot y))-N(Nx\cdot Dy+x\cdot NDy-N(x\cdot Dy))\\
&=&N(DNx\cdot y-Nx\cdot Dy+Dx\cdot Ny-x\cdot DNy-N(Dx\cdot y-x\cdot Dy))\\
&=&N([Nx,y]_D+[x,Ny]_D-N[x,y]_D),
\end{eqnarray*}
which implies that $N$ is a Nijenhuis operator on the Lie algebra $(\g,[\cdot,\cdot]_D)$. By Theorem $\ref{pro:NijtoNij}$, $N$ is a Nijenhuis operator on the $3$-Lie algebra $(\g,\llbracket \cdot,\cdot,\cdot\rrbracket)$.\qed\vspace{3mm}

Let $(\g,\cdot)$ be a commutative associative algebra.
For $x_i,~y_i,~z_i\in \g,~i=1,~2,~3$, denote by
\begin{eqnarray*}
\begin{vmatrix}\vec{x}&\vec{y}&\vec{z}\end{vmatrix}&=&\begin{vmatrix}x_1& y_1&z_1\\ x_2&y_2&z_2\\ x_3&y_3&z_3\end{vmatrix}\\
&=&x_1\cdot(y_2\cdot z_3-y_3\cdot z_2)-x_2\cdot(y_1\cdot z_3-y_3\cdot z_1)+x_3\cdot(y_1\cdot z_2-y_2\cdot z_1),
\end{eqnarray*}
where $\vec{x}$, $\vec{y}$ and $\vec{z}$ denote the column vectors.
\begin{lem}\label{lem:calcu det}
Let $N$ be a Nijenhuis operator on  a commutative associative algebra $(\g,\cdot)$ and $N(\vec{x}),N(\vec{y}),N(\vec{z})$ denote the images of the column vectors. Then we have
\begin{eqnarray*}
\begin{vmatrix}N(\vec{x})&N(\vec{y})&N(\vec{z})\end{vmatrix}&=&N\left(\begin{vmatrix}N(\vec{x})&N(\vec{y})&\vec{z}\end{vmatrix}+c.p.\right)-N^2\left(\begin{vmatrix}N(\vec{x})&\vec{y}&\vec{z}\end{vmatrix}+c.p.\right)\\
&&+N^3\left(\begin{vmatrix}\vec{x}&\vec{y}&\vec{z}\end{vmatrix}\right).
\end{eqnarray*}
\end{lem}
\pf Since $N$ is a Nijenhuis operator on $(\g,\cdot)$, we have
 \begin{eqnarray*}
\begin{vmatrix}N(\vec{x})&N(\vec{y})&N(\vec{z})\end{vmatrix}&=&\sum_{\sigma\in S_3}\sgn(\sigma)N(x_{\sigma(1)})N(y_{\sigma(2)})N(z_{\sigma(3)})\\
&=&N\big(\sum_{\sigma\in S_3}\sgn(\sigma)N(x_{\sigma(1)})N(y_{\sigma(2)})z_{\sigma(3)}+c.p.\big)\\&&-N^2\big(\sum_{\sigma\in S_3}\sgn(\sigma)N(x_{\sigma(1)})y_{\sigma(2)}z_{\sigma(3)}+c.p.\big)\\&&+N^3\big(\sum_{\sigma\in S_3}\sgn(\sigma)x_{\sigma(1)}y_{\sigma(2)}z_{\sigma(3)}\big)\\
&=&N\left(\begin{vmatrix}N(\vec{x})&N(\vec{y})&\vec{z}\end{vmatrix}+c.p.\right)-N^2\left(\begin{vmatrix}N(\vec{x})&\vec{y}&\vec{z}\end{vmatrix}+c.p.\right)\\
&&+N^3\left(\begin{vmatrix}\vec{x}&\vec{y}&\vec{z}\end{vmatrix}\right).
\end{eqnarray*}
The proof is finished. \qed

\begin{lem}\label{lem:2derivation}{\rm(\cite{BaiGuo})}
Let $(\g,\cdot)$ be a commutative associative algebra, $D_1,D_2\in\Der(\g)$ satisfying $D_1D_2=D_2D_1$. Then
$(\g,\llbracket \cdot,\cdot,\cdot\rrbracket)$ is a $3$-Lie algebra, where the bracket is given by
\begin{eqnarray}\label{eq:Rota-Baxter 2}
\llbracket x,y,z\rrbracket\triangleq \begin{vmatrix}x& y&z\\ D_1(x)&D_1(y)&D_1(z)\\ D_2(x)&D_2(y)&D_2(z)\end{vmatrix}, \quad \forall x,y,z\in\g.
\end{eqnarray}
\end{lem}

\begin{pro}\label{pro:ex2}
With the same assumptions as Lemma \ref{lem:2derivation}. Let  $N$ be a Nijenhuis operator on $(\g,\cdot)$   satisfying   $ND_1=D_1N,~  ND_2=D_2N$. Then $N$ is a Nijenhuis operator on the $3$-Lie algebra $(\g,\llbracket \cdot,\cdot,\cdot\rrbracket)$, where the bracket is given by Eq. $\eqref{eq:Rota-Baxter 2}$.
\end{pro}
\pf Since $ND_1=D_1N,\  ND_2=D_2N$, by Lemma \ref{lem:calcu det}, we have
\begin{eqnarray*}
\llbracket Nx,Ny,Nz\rrbracket&=& \begin{vmatrix}Nx& Ny&Nz\\ D_1(Nx)&D_1(Ny)&D_1(Nz)\\ D_2(Nx)&D_2(Ny)&D_2(Nz)\end{vmatrix}\\
&=&\begin{vmatrix}Nx& Ny&Nz\\ ND_1(x)&ND_1(y)&ND_1(z)\\ ND_2(x)&ND_2(y)&ND_2(z)\end{vmatrix}\\
&=&N\left(\begin{vmatrix}Nx& Ny&z\\ ND_1(x)&ND_1(y)&D_1(z)\\ ND_2(x)&ND_2(y)&D_2(z)\end{vmatrix}+c.p.\right)\\
&&-N^2\left(\begin{vmatrix}Nx& y&z\\ ND_1(x)&D_1(y)&D_1(z)\\ ND_2(x)&D_2(y)&D_2(z)\end{vmatrix}+c.p.\right)+N^3\left(\begin{vmatrix}x& y&z\\ D_1(x)&D_1(y)&D_1(z)\\ D_2(x)&D_2(y)&D_2(z)\end{vmatrix}\right)\\
&=&N\left(\begin{vmatrix}Nx& Ny&z\\ D_1(Nx)&D_1(Ny)&D_1(z)\\ D_2(Nx)&D_2(Ny)&D_2(z)\end{vmatrix}+c.p.\right)\\
&&-N^2\left(\begin{vmatrix}Nx& y&z\\ D_1(Nx)&D_1(y)&D_1(z)\\ D_2(Nx)&D_2(y)&D_2(z)\end{vmatrix}+c.p.\right)+N^3\left(\begin{vmatrix}x& y&z\\ D_1(x)&D_1(y)&D_1(z)\\ D_2(x)&D_2(y)&D_2(z)\end{vmatrix}\right)\\
&=&N(\llbracket Nx,Ny,z\rrbracket+c.p.)-N^2(\llbracket Nx,y,z\rrbracket+c.p.)+N^3(\llbracket x,y,z\rrbracket)\\
&=&N(\llbracket x,y,z\rrbracket_N^2).
\end{eqnarray*}
Thus, $N$ is a Nijenhuis operator on the $3$-Lie algebra  $(\g,\llbracket \cdot,\cdot,\cdot\rrbracket)$.\qed

\begin{lem}\label{lem:3derivation}{\rm(\cite{BaiGuo})}
Let $(\g,\cdot)$ be a commutative associative algebra. Let $D_i\in\Der(\g)$ such that $D_iD_j=D_jD_i,\ i,j=1,2,3$. Then
$(\g,\llbracket \cdot,\cdot,\cdot\rrbracket)$ is a $3$-Lie algebra, where the bracket is given by
\begin{eqnarray}\label{eq:Rota-Baxter 3}
\llbracket x,y,z\rrbracket:= \begin{vmatrix}D_1(x)&D_1(y)&D_1(z)\\ D_2(x)&D_2(y)&D_2(z)\\ D_3(x)&D_3(y)&D_3(z)\end{vmatrix},\quad \forall x,y,z\in\g.
\end{eqnarray}
\end{lem}

\begin{pro}
With the same assumptions as Lemma \ref{lem:3derivation}. Let $N$ be a Nijenhuis operator on $(\g,\cdot)$ satisfying $ND_i=D_iN,\ i,j=1,2,3$. Then $N$ is a Nijenhuis operator on the $3$-Lie algebra $(\g,\llbracket \cdot,\cdot,\cdot\rrbracket)$, where the bracket is given by Eq. $\eqref{eq:Rota-Baxter 3}$.
\end{pro}
\pf The proof is similar to the proof of Proposition \ref{pro:ex2}. We omit details.\qed

\emptycomment{\begin{cor}
With the same assumptions in Lemma $\ref{lem:RBDer}$, if $N$ is invertible, then $N^{-1}$
 is also a Nijenhuis operator.
\end{cor}
\pf For all $x_1=N^{-1}y_1,x_2=N^{-1}y_2,x_3=N^{-1}y_3$, by (\ref{RB}), we have
\begin{eqnarray*}
N^{-1}[y_1,y_2,y_3]&=&N^{-1}[Nx_1,Nx_2,Nx_3]\\
&=&N^{-1}N([Nx_1,Nx_2,x_3]+[Nx_1,x_2,Nx_3]+[x_1,Nx_2,Nx_3])\\
&=&[y_1,y_2,N^{-1}y_3]+[y_1,N^{-1}y_2,y_3]+[N^{-1}y_1,y_2,y_3].
\end{eqnarray*}
And, by (\ref{Deri}), we have
\begin{eqnarray*}
N[N^{-1}y_1,N^{-1}y_2,N^{-1}y_3]=[y_1,N^{-1}y_2,N^{-1}y_3]+[N^{-1}y_1,y_2,N^{-1}y_3]+[N^{-1}y_1,N^{-1}y_2,y_3],
\end{eqnarray*}
which is equivalent to
\begin{eqnarray*}
&&[N^{-1}y_1,N^{-1}y_2,N^{-1}y_3]\\
&=&N^{-1}([y_1,N^{-1}y_2,N^{-1}y_3]+[N^{-1}y_1,y_2,N^{-1}y_3]+[N^{-1}y_1,N^{-1}y_2,y_3]).
\end{eqnarray*}
By Proposition \ref{lem:RBDer}, the corollary follows immediately.\qed\vspace{3mm}

 Assume that $(\g,[\cdot,\cdot,\cdot])$ is a $3$-Lie algebra with a Nijenhuis operator $N$.  By Corollary \ref{cor:n-LA},  $(\g,[\cdot,\cdot,\cdot]_N^2)$ is also a $3$-Lie algebra.

\begin{thm}\label{pro:RD-relation}
Let $N$ be a Nijenhuis operator on the $3$-Lie algebra$(\g,[\cdot,\cdot,\cdot])$. If $N$ is a also derivation on the $3$-Lie algebra $(\g,[\cdot,\cdot,\cdot])$, then $N$ is a Nijenhuis operator on the $3$-Lie algebra $(\g,[\cdot,\cdot,\cdot]_N^2)$.
\end{thm}
\pf Since $N$ is a derivation on the $3$-Lie algebra $(\g,[\cdot,\cdot,\cdot])$, we have
 \begin{eqnarray*}
 N[x,y,z]_N^2&=&N[Nx,Ny,z]+N[Nx,y,Nz]+N[x,Ny,Nz]\\
 &=&[N^2x,Ny,z]+[Nx,N^2y,z]+[Nx,Ny,NZ]\\
 &&+[N^2x,y,Nz]+[Nx,Ny,Nz]+[Nx,y,N^2Z]\\
 &&+[x,N^2y,Nz]+[x,Ny,N^2z]++[Nx,Ny,NZ]\\
 &=&[N^2x,Ny,z]+[N^2x,y,Nz]+[Nx,Ny,NZ]\\
 &&+[x,Ny,N^2z]+[Nx,Ny,Nz]+[Nx,y,N^2Z]\\
 &&+[x,N^2y,Nz]+[Nx,N^2y,z]+[Nx,Ny,NZ]\\
 &=&[Nx,y,z]_N^2+[x,Ny,z]_N^2+[x,y,Nz]_N^2,
 \end{eqnarray*}
which implies that $N$ is a derivation on the $3$-Lie algebra $(\g,[\cdot,\cdot,\cdot]_N^2)$.

By \eqref{eq:3Nijenhuis},
we have $$[Nx,Ny,Nz]=N[x,y,z]_N^2.$$
Therefore, we have
\begin{eqnarray*}
[Nx,Ny,Nz]_N^2&=&[N^2x,N^2y,Nz]+[N^2x,Ny,N^2z]+[Nx,N^2y,N^2z]\\
&=& N([Nx,Ny,z]_N^2+[Nx,y,Nz]_N^2+[x,Ny,Nz]_N^2),
\end{eqnarray*}
which implies $N$ is a Rota-Baxter operator on the $3$-Lie algebra $(\g,[\cdot,\cdot,\cdot]_N^2)$.

By Lemma \ref{lem:RBDer}, $N$ is a Nijenhuis operator on the $3$-Lie algebra $(\g,[\cdot,\cdot,\cdot]_N^2)$.\qed

Now we will introduce some notations which will be used below.
\begin{eqnarray*}
[x,y,z]_{(i+1)}^N&=&[Nx,Ny,z]_{i}^N+[x,Ny,Nz]_{i}^N+[Nx,y,Nz]_{i}^N,\quad i\in\Numb.
\end{eqnarray*}
For $i=0$, we define
\begin{equation}
[x,y,z]_{0}^N=[x,y,z].
\end{equation}
Therefore for $i=1$, we have
\begin{equation}
[x,y,z]_{1}^N=[x,y,z]_N^2.
\end{equation}

\begin{thm}
Assume that $(\g,[\cdot,\cdot,\cdot])$ is a $3$-Lie algebra with a Nijenhuis operator $N$. If $N$ is a derivation of a $3$-Lie algebra $(\g,[\cdot,\cdot,\cdot])$, then we have
 \begin{itemize}
     \item[\rm(i)] $(\g,[\cdot,\cdot,\cdot]_{i}^N)$ is a $3$-Lie algebra.
     \item[\rm(ii)] $N$ is a derivation on the $3$-Lie algebra $(\g,[\cdot,\cdot,\cdot]_{i}^N)$.
     \item[\rm(iii)] $N$ is a Rota-Baxter operator on the $3$-Lie algebra $(\g,[\cdot,\cdot,\cdot]_{i}^N)$.
     \item[\rm(iv)] $N$ is a Nijenhuis operator on the $3$-Lie algebra $(\g,[\cdot,\cdot,\cdot]_{i}^N)$.
   \end{itemize}
\end{thm}
\pf By induction on $i$, and using the same methods in Proposition \ref{pro:RD-relation}, the theorem follows directly.\qed}

%Let $(\g,[\cdot,\cdot,\cdot])$ be a $3$-Lie algebra, denote $\g^1=[\g,\g,\g]$.

%\begin{lem}
%Let $\g$ be a $4$-dimensional complex $3$-Lie algebra with a basis $\{e_1,e_2,e_3,e_4\}$. Then $\g$ is isomorphic to one of the following $3$-Lie algebras:
 %\begin{itemize}

     %\item[\rm(a)] If dim $\g^1$ =\rm0, then $\g$ is an abelian $3$-Lie algebra.
     %\item[\rm(b)] If dim $\g^1$ =\rm1, then there are two cases:
    % \begin{eqnarray}(b_1)~~[e_2,e_3,e_4]=e_1,\quad   (b_2)~~[e_1,e_2,e_3]=e_1.\end{eqnarray}
   %  \item[\rm(c)]  If dim $\g^1$ =2, then there are three cases:
  %\begin{eqnarray}
%(c_1)~~\left\{\begin{array}{ccc}[e_2,e_3,e_4]&=&e_1\\

 %[e_1,e_3,e_4]&=&e_2,\end{array}\right.\quad (c_2)~~\left\{\begin{array}{ccc}[e_2,e_3,e_4]&=&\alpha e_1+e_2,\\

 %[e_1,e_3,e_4]&=&e_2,\end{array}\right.\quad \alpha\neq 0,
%\end{eqnarray}
 %\begin{eqnarray}
%(c_1)~~\left\{\begin{array}{ccc}[e_1,e_3,e_4]&=&e_1\\

 %[e_2,e_3,e_4]&=&e_2.\end{array}\right.
 %\end{eqnarray}
  %   \item[\rm(d)] If dim $\g^1$ =\rm3, then there is only one case:
   %  \begin{eqnarray}(d)~~[e_2,e_3,e_4]=e_1,~~[e_1,e_2,e_4]=e_3,~~[e_1,e_3,e_4]=e_2.\end{eqnarray}
    % \item[\rm(e)] If dim $\g^1$ =\rm4, then there is only one case:
     %\begin{eqnarray}(e)~~[e_2,e_3,e_4]=e_1,~~[e_1,e_2,e_4]=e_3,~~[e_1,e_3,e_4]=e_2,~~[e_1,e_2,e_3]=e_4.\end{eqnarray}
   %\end{itemize}
%\end{lem}}
  \subsection{Constructions of Nijenhuis operators on $3$-Lie algebras from Rota-Baxter operators and derivations and some explicit examples}

Recall that a {\bf Rota-Baxter operator} (of weight $0$) on a $3$-Lie
 algebra $(\g,[\cdot,\cdot,\cdot])$ is a linear map $P:\g\longrightarrow\g$ such that
\begin{equation}\label{RB}
[Px,Py,Pz]=P([Px,Py,z]+[Px,y,Pz]+[x,Py,Pz]),\quad \forall x,y,z\in\g,
\end{equation}
and a {\bf derivation} on a $3$-Lie
 algebra $(\g,[\cdot,\cdot,\cdot])$ is a linear map $D:\g\longrightarrow\g$ such that
\begin{equation}\label{Deri}
D[x,y,z]=[Dx,y,z]+[x,Dy,z]+[x,y,Dz],\quad \forall x,y,z\in\g.
\end{equation}
We denote the sets of Rota-Baxter operators (of weight 0) and derivations of a 3-Lie algebra $\g$ by
${\rm RB}(\g)$ and ${\rm Der}(\g)$ respectively. Note that ${\rm Der}(\g)$ is a vector space, where
${\rm RB}(\g)$ is not a vector space (it is only a set!).

The following conclusion is straightforward but very important  for constructing Nijenhuis operators.
\begin{lem}\label{lem:RBDer}
Let $(\g,[\cdot,\cdot,\cdot])$ be a $3$-Lie
 algebra.
 If a linear transformation $N$ is a derivation, then $N$ is a Nijenhuis operator if and only if
 $N$ is a Rota-Baxter operator (of weight 0) on $\g$. In particular,
 if a linear transformation $N\in {\rm RB}(\g)\cap {\rm Der}(\g)$,
 then $N$ is a Nijenhuis operator.
\end{lem}

\begin{ex}{\rm
Let $\g$ be the $4$-dimensional simple complex $3$-Lie algebra  given by
 \begin{eqnarray}
 [e_2,e_3,e_4]=e_1,~~[e_1,e_2,e_4]=e_3,~~[e_1,e_3,e_4]=e_2,~~[e_1,e_2,e_3]=e_4,
 \end{eqnarray}
 where $\{e_1,e_2,e_3,e_4\}$ is a basis of $\g$.
Then by direct computations, we have
$$
\Der(\g)=\Big\{\begin{bmatrix}0& a&b&c\\ a&0&d&e\\ -b&d&0&f\\c&-e&f&0\end{bmatrix}\Big| a,b,c,d,e,f\in\Comp\Big\}.
$$
Let $N\in\Der(\g)$. Then we have
$$
[e_2,e_3,e_4]_N^1=0,~~[e_1,e_2,e_4]_N^1=0,~~[e_1,e_3,e_4]_N^1=0,~~[e_1,e_2,e_3]_N^1=0,
$$
  In our convention, $Ne_i=N_i^je_j$. Furthermore, we have
\begin{eqnarray*}
[e_2,e_3,e_4]_N^2&=&(-d^2+e^2-f^2)e_1+(bd-ce)e_2+(ad+cf)e_3+(ae+bf)e_4,\\
{[e_1,e_2,e_4]}_N^2&=&(ad+cf)e_1-(ba+ef)e_2-(a^2+c^2-e^2)e_3-(bc-de)e_4,\\
{[e_1,e_3,e_4]}_N^2&=&(ce-bd)e_1+(b^2-c^2-f^2)e_2+(ba+ef)e_3+(ca+df)e_4,\\
{[e_1,e_2,e_3]}_N^2&=&-(ae+bf)e_1+(ac+df)e_2+(bc-de)e_3+(b^2-a^2-d^2)e_4
\end{eqnarray*}
It is straightforward to deduce that $N[x,y,z]_N^2=[Nx,Ny,Nz]$ for all $x,y,z\in \g$.
  Therefore,  $N$ is a Rota-Baxter operator, i.e. ${\rm Der}(\g)\subset {\rm RB}(\g)$. Thus, any $N\in {\rm Der} (\g)$ is a Nijenhuis operator.

  By Theorem $\ref{th:polnij}$, $N^2$ is also a Nijenhuis operator. However, it is straightforward to deduce that   $N^2$ is neither a derivation nor a Rota-Baxter operator any more.

\emptycomment{
On the other hand, we can get
\begin{eqnarray}
N^2=\begin{bmatrix}-b^2+c^2+a^2& bd-ce&ad+cf&ae+bf\\ -bd+ce&a^2+d^2-e^2&ab+ef&dc+df\\ ad+cf&-ab-ef&-b^2+d^2+f^2&-bc+de\\-ae-bf&ac+df&bc-de&c^2-e^2+f^2\end{bmatrix}.
\end{eqnarray}
Then by direct computations, we have
\begin{eqnarray*}
[e_2,e_3,e_4]_{N^2}^1&=&(2 d^2-2 e^2+2 f^2)e_1+(-2 b d+2 c e)e_2+(-2 a d-2 c f)e_3+(-2 a e-2 b f)e_4,\\
{[e_1,e_2,e_4]}_{N^2}^1&=&(-2 a d-2 c f)e_1+(2 a b+2 e f)e_2+(2 a^2+2 c^2-2 e^2)e_3+(2 b c-2 d e)e_4,\\
{[e_1,e_3,e_4]}_{N^2}^1&=&(2 b d-2 c e)e_1+(-2 b^2+2 c^2+2 f^2)e_2+(-2 a b-2 e f)e_3+(-2 a c-2 d f)e_4,\\
{[e_1,e_2,e_3]}_{N^2}^1&=&(2 a e+2 b f)e_1+(-2 a c-2 d f)e_2+(-2 b c+2 d e)e_3+(2 a^2-2 b^2+2 d^2)e_4.
\end{eqnarray*}
From this we can see that
and
\begin{eqnarray*}
[e_2,e_3,e_4]_{N^2}^2&=&\Big(2 c^2 d^2+d^4-2 b c d e-c^2 e^2-2 d^2 e^2+e^4+c^2 f^2+2 d^2 f^2-2 e^2 f^2\\&&+f^4+a (-2 c d f+2 b e f)-b^2 (d^2-2 e^2+f^2)+a^2 (d^2-e^2+2 f^2)\Big)e_1\\
&&+\Big(-(b d-c e) (a^2-b^2+c^2+d^2-e^2+f^2)\Big)e_2\\
&&+\Big(-(a d+c f) (a^2-b^2+c^2+d^2-e^2+f^2)\Big)e_3\\
&&+\Big(-(a e+b f) (a^2-b^2+c^2+d^2-e^2+f^2)\Big)e_4,\\
{[e_1,e_2,e_4]}_{N^2}^2&=&\Big(-(a d+c f) (a^2-b^2+c^2+d^2-e^2+f^2)\Big)e_1\\
&&+\Big((a b+e f) (a^2-b^2+c^2+d^2-e^2+f^2)\Big)e_2\\
&&+\Big(a^4+c^4+2 c^2 d^2-2 b c d e-2 c^2 e^2-d^2 e^2+e^4-b^2 (c^2-2 e^2)\\&&+c^2 f^2-e^2 f^2+a (-2 c d f+2 b e f)+a^2 (-b^2+2 c^2+d^2-2 e^2+2 f^2)\Big)e_3\\
&&+\Big((b c-d e) (a^2-b^2+c^2+d^2-e^2+f^2)\Big)e_4,\\
{[e_1,e_3,e_4]}_{N^2}^2&=&\Big((b d-c e) (a^2-b^2+c^2+d^2-e^2+f^2)\Big)e_1\\
&&+\Big(b^4+c^4+2 c^2 d^2-2 b c d e-c^2 e^2+2 c^2 f^2+d^2 f^2-e^2 f^2+f^4\\
&&+a (-2 c d f+2 b e f)+a^2 (-b^2+c^2+2 f^2)-b^2 (2 c^2+d^2-2 e^2+2 f^2)\Big)e_2\\
&&+\Big(-(a b+e f) (a^2-b^2+c^2+d^2-e^2+f^2)\Big)e_3\\
&&+\Big(-(a c+d f) (a^2-b^2+c^2+d^2-e^2+f^2)\Big)e_4,\\
{[e_1,e_2,e_3]}_{N^2}^2&=&\Big((a e+b f) (a^2-b^2+c^2+d^2-e^2+f^2)\Big)e_1\\&&-\Big((a c+d f) (a^2-b^2+c^2+d^2-e^2+f^2)\Big)e_2\\
&&+\Big((b c-d e) (a^2-b^2+c^2+d^2-e^2+f^2)\Big)e_3\\
&&+\Big(a^4+b^4-2 b c d e+a (-2 c d f+2 b e f)-b^2 (c^2+2 d^2-2 e^2+f^2)\\
&&+d^2 (2 c^2+d^2-e^2+f^2)+a^2 (-2 b^2+c^2+2 d^2-e^2+2 f^2)\Big)e_4.
\end{eqnarray*}
It is straightforward to deduce that $N^2[x,y,z]_N^2=[N^2x,N^2y,N^2z]$. Thus, $N^2$ is a Nijenhuis operator.
}}
\end{ex}

\begin{ex}{\rm
Let $\g$ be the $4$-dimensional complex $3$-Lie algebra  given by
 \begin{eqnarray}
 [e_2,e_3,e_4]=e_1,~~[e_1,e_2,e_4]=e_3,~~[e_1,e_3,e_4]=e_2,
 \end{eqnarray}
 where $\{e_1,e_2,e_3,e_4\}$ is a basis of $\g$. Then we have
  $$
 \Der(\g)=\Big\{\begin{bmatrix}h& a&b&0\\ a&h&c&0\\ -b&c&h&0\\d&e&f&-h\end{bmatrix}\Big| a,b,c,d,e,f,h\in\Comp\Big\}.
 $$
In general, a derivation $D\in {\rm Der}(\g)$ might not be a Rota-Baxter operator (of weight 0) of $\g$ any more. On the other hand, denote by $T_1(\g)\subset \Der(\g)$ and  $T_2(\g)\subset \Der(\g)$ respectively two   subspaces of $\Der(\g)$ given by
$$
T_1(\g)=\Big\{\begin{bmatrix}0& a&b&0\\ a&0&c&0\\ -b&c&0&0\\d&e&f&0\end{bmatrix}\Big| a,b,c,d,e,f\in\Comp\Big\},
$$
and
$$
T_2(\g)=\Big\{\begin{bmatrix}a& 0&0&0\\ 0&a&0&0\\ 0&0&a&0\\b&c&d&-a\end{bmatrix}\Big|a,b,c,d\in\Comp\Big\}.
$$
Then ${\rm Der} (\g)=T_1(\g)+ T_2(\g)$. Furthermore, we can deduce that a derivation $D\in {\rm Der}(\g)$ is a Rota-Baxter operator (of weight 0) of $\g$ if and only if $D\in T_1(\g)$, or $D\in T_2(\g)$. Thus, a derivation $D\in {\rm Der}(\g)$ is a Nijenhuis operator if and only if $D\in T_1(\g)$, or $D\in T_2(\g)$.
\emptycomment{
For any $N\in T_1(\g)$ and $\tilde{N}\in T_2(\g)$, we have
\emptycomment{\begin{eqnarray*}
[e_2,e_3,e_4]_N^1&=&0,~~[e_1,e_2,e_4]_N^1=0,~~[e_1,e_3,e_4]_N^1=0,~~[e_1,e_2,e_3]_N^1=0,\\
{[e_2,e_3,e_4]}_{\tilde{N}}^1&=&0,~~{[e_1,e_2,e_4]}_{\tilde{N}}^1=0,~~{[e_1,e_3,e_4]}_{\tilde{N}}^1=0,~~{[e_1,e_2,e_3]}_{\tilde{N}}^1=0,
\end{eqnarray*}
and}
\begin{eqnarray*}
[e_1,e_2,e_3]_N^2&=&0,\\
{[e_2,e_3,e_4]}_N^2&=&-(c^2)e_1+(bc)e_2+(ac)e_3,\\
{[e_1,e_2,e_4]}_N^2&=&(ac)e_1-(ba)e_2-(a^2)e_3,\\
{[e_1,e_3,e_4]}_N^2&=&-(bc)e_1+(b^2)e_2+(ba)e_3,\\
{[e_1,e_2,e_3]}_{\tilde{N}}^2&=&0,\\
{[e_2,e_3,e_4]}_{\tilde{N}}^2&=&-(a^2)e_1,\\
{[e_1,e_2,e_4]}_{\tilde{N}}^2&=&-(a^2)e_3,\\
{[e_1,e_3,e_4]}_{\tilde{N}}^2&=&-(a^2)e_2.
\end{eqnarray*}
}}
\end{ex}

\smallskip

\noindent {\bf Acknowledgements: } This research  is supported by
NSF of China (11471139, 11271202, 11221091, 11425104), SRFDP
(20120031110022) and NSF of Jilin Province (20140520054JH).

\section*{Appendix}
{\bf The proof of Lemma \ref{lem:impotant relation}}:
Fix $\alpha_1=1,\alpha_2=1,\cdots,\alpha_{n-1}=1$ and prove Eq.~$\eqref{eq:impotant relation}$ for arbitrary $\alpha_n>0$. For $\alpha_n=1$, the formula is just Eq. $\eqref{eq:Nijenhuis all}$. Now assume that Eq.~ $\eqref{eq:impotant relation}$ holds for $\alpha_n=\beta_n$. By Eq. $\eqref{eq:Nijenhuis all}$, for $\alpha_n=\beta_{n}+1$, we get

 \begin{eqnarray*}
 &&[Nx_1,Nx_2,\cdots,Nx_{n-1},N^{\beta_n+1}x_n]+\sum_{p=1}^n\sum_{\sigma}(-1)^{\frac{p(p-1)}{2}+\sum_{j=1}^p{\sigma(j)}}N^{\sum_{j=1}^p\alpha_{\sigma(j)}}\\&&
 [x_{\sigma(1)},x_{\sigma(2)},\cdots,x_{\sigma(p)},N^{\alpha_{\sigma(p+1)}}x_{\sigma(p+1)},\cdots,N^{\alpha_{\sigma(n)}}x_{\sigma(n)}] \\
 &=&\sum_{p=1}^n\sum_{\sigma}\big(-(-1)^{\frac{p(p-1)}{2}+\sum_{j=1}^p\sigma(j)}N^p[x_{\sigma(1)},x_{\sigma(2)},\cdots,x_{\sigma(p)},Nx_{\sigma(p+1)},\cdots,Nx_{\sigma(n)}]
 \\&&+(-1)^{\frac{p(p-1)}{2}+\sum_{j=1}^p\sigma(j)}N^{\sum_{j=1}^p\alpha_{\sigma(j)}}\\&&[x_{\sigma(1)},x_{\sigma(2)},\cdots,x_{\sigma(p)},N^{\alpha_{\sigma(p+1)}}x_{\sigma(p+1)},
 \cdots,N^{\alpha_{\sigma(n)}}x_{\sigma(n)}]\big)
 \end{eqnarray*}
  \begin{eqnarray*}
 &=&\sum_{p=1}^n\big(-\sum_{\sigma}(-1)^{\frac{p(p-1)}{2}+\sum_{j=1}^{p-1}\sigma(j)+n}N^p[x_{\sigma(1)},x_{\sigma(2)},\cdots,N^{\beta_n}x_{n},Nx_{\sigma(p+1)},\cdots,Nx_{\sigma(n)}]\\
 &&-\sum_{\sigma}(-1)^{\frac{p(p-1)}{2}+\sum_{j=1}^p\sigma(j)}N^p[x_{\sigma(1)},x_{\sigma(2)},\cdots,x_{\sigma(p)},Nx_{\sigma(p+1)},\cdots,N^{\beta_n+1}x_{n}]\\
 &&+\sum_{\sigma}(-1)^{\frac{p(p-1)}{2}+\sum_{j=1}^{p-1}\sigma(j)+n}N^{\sum_{j=1}^{p-1}\alpha_{\sigma(j)}+\beta_n+1}\\&&[x_{\sigma(1)},x_{\sigma(2)},\cdots,x_{n},N^{\alpha_{\sigma(p+1)}}x_{\sigma(p+1)},\cdots,N^{\alpha_{\sigma(n)}}x_{\sigma(n)}]\\
 &&+\sum_{\sigma}(-1)^{\frac{p(p-1)}{2}+\sum_{j=1}^p\sigma(j)}N^p[x_{\sigma(1)},x_{\sigma(2)},\cdots,x_{\sigma(p)},Nx_{\sigma(p+1)},\cdots,N^{\beta_n+1}x_{n}]\big)\\
 &=&-\sum_{p=1}^n\sum_{\sigma}(-1)^{\frac{p(p-1)}{2}+\sum_{j=1}^{p-1}\sigma(j)+n}N^p[x_{\sigma(1)},x_{\sigma(2)},\cdots,N^{\beta_n}x_{n},Nx_{\sigma(p+1)},\cdots,Nx_{\sigma(n)}]\\
 &&+\sum_{p=1}^n\sum_{\sigma}(-1)^{\frac{p(p-1)}{2}+\sum_{j=1}^{p-1}\sigma(j)+n}N^{\sum_{j=1}^{p-1}\alpha_{\sigma(j)}+\beta_n+1}\\&&[x_{\sigma(1)},x_{\sigma(2)},\cdots,x_{n},N^{\alpha_{\sigma(p+1)}}x_{\sigma(p+1)},\cdots,N^{\alpha_{\sigma(n)}}x_{\sigma(n)}]\\
  &=&\sum_{p=1}^n\sum_{\sigma}(-1)^{\frac{(p-1)(p-2)}{2}+\sum_{j=1}^{p-1}\sigma(j)}N^p\\&&[x_{\sigma(1)},x_{\sigma(2)},\cdots,x_{\sigma(p-1)},Nx_{\sigma(p+1)},\cdots,Nx_{\sigma(n)},N^{\beta_n}x_{n}]\\
 &&+\sum_{p=1}^n\sum_{\sigma}(-1)^{\frac{p(p-1)}{2}+\sum_{j=1}^{p-1}\sigma(j)+n}N^{\sum_{j=1}^{p-1}\alpha_{\sigma(j)}+\beta_n+1}\\&&[x_{\sigma(1)},x_{\sigma(2)},\cdots,x_{n},N^{\alpha_{\sigma(p+1)}}x_{\sigma(p+1)},\cdots,N^{\alpha_{\sigma(n)}}x_{\sigma(n)}]\\
 &=&N\Big([Nx_1,Nx_2,\cdots,Nx_{n-1},N^{\beta_n}x_n]+\sum_{p=1}^n\sum_{\sigma}(-1)^{\frac{p(p-1)}{2}+\sum_{j=1}^p\sigma(j)}N^{\sum_{j=1}^p\alpha_{\sigma(j)}}
 \\&&[x_{\sigma(1)},x_{\sigma(2)},\cdots,x_{\sigma(p)},N^{\alpha_{\sigma(p+1)}}x_{\sigma(p+1)},\cdots,N^{\alpha_{\sigma(n)}}x_{\sigma(n)}]\Big),
 \end{eqnarray*}
which implies that Eq.$~\eqref{eq:Nijenhuis all}$ holds for $\alpha_1=1,\alpha_2=1,\cdots,\alpha_{n-1}=1$ and arbitrary positive integer $\alpha_n$.

 \emptycomment{Now let $\alpha_1=1,\alpha_2=1,\cdots,\alpha_{n-2}=1,\alpha_{n-1}=\beta_{n-1}+1,\alpha_n$. Applying the Eq.$~\eqref{eq:Nijenhuis all}$ for $\alpha_1=1,\alpha_2=1,\cdots,\alpha_{n-1}=1,\alpha_n$ to the element $N^{\beta_{n-1}}x_{n-1}$ instead of the element $x_{n-1}$ and let $y_1=x_1,\cdots,y_{n-1}=N^{\beta_{n-1}}x_{n-1},y_n=x_n$, by Eq.$~\eqref{eq:Nijenhuis all}$, we have
 \begin{eqnarray*}
 &&[Ny_1,Ny_2,\cdots,Ny_{n-1},N^{\alpha_n}y_n]+\sum_{p=1}^n\sum_{\sigma}(-1)^{\frac{p(p-1)}{2}+\sum_{j=1}^p\sigma(j)}N^{\sum_{j=1}^p\alpha_{\sigma(j)}}\\&&
 [x_{\sigma(1)},x_{\sigma(2)},\cdots,x_{\sigma(p)},N^{\alpha_{\sigma(p+1)}}x_{\sigma(p+1)},\cdots,N^{\alpha_{\sigma(n)}}x_{\sigma(n)}]\\
 &=&-\sum_{p=1}^n\sum_{\sigma}(-1)^{\frac{p(p-1)}{2}+\sum_{j=1}^p\sigma(j)}N^{\sum_{j=1}^p\alpha_{\sigma(j)}}\\&&[y_{\sigma(1)},y_{\sigma(2)},\cdots,y_{\sigma(p)},N^{\alpha_{\sigma(p+1)}}y_{\sigma(p+1)},\cdots,
 N^{\alpha_{\sigma(n)}}y_{\sigma(n)}]\\&&+\sum_{p=1}^n\sum_{\sigma}(-1)^{\frac{p(p-1)}{2}+\sum_{j=1}^p\sigma(j)}N^{\sum_{j=1}^p\alpha_{\sigma(j)}}
 \\&&[x_{\sigma(1)},x_{\sigma(2)},\cdots,x_{\sigma(p)},N^{\alpha_{\sigma(p+1)}}x_{\sigma(p+1)},\cdots,N^{\alpha_{\sigma(n)}}x_{\sigma(n)}]\\
 &=&-\sum_{p=1}^n\Big(\sum_{\sigma}(-1)^{\frac{p(p-1)}{2}+\sum_{j=1}^{p-1}\sigma(j)+n-1}N^{p}\\&&[x_{\sigma(1)},x_{\sigma(2)},\cdots,N^{\beta_{n-1}}x_{n-1},N^{\alpha_{\sigma(p+1)}}x_{\sigma(p+1)},\cdots,
 N^{\alpha_{n}}x_{n}]\\&&+\sum_{\sigma}(-1)^{\frac{p(p-1)}{2}+\sum_{j=1}^{p-2}\sigma(j)+2n-1}N^{p+\alpha_n-1}\\&&[x_{\sigma(1)},x_{\sigma(2)},\cdots,N^{\beta_{n-1}}x_{n-1},x_{n},N^{\alpha_{\sigma(p+1)}}y_{\sigma(p+1)},\cdots,
 N^{\alpha_{\sigma(n)}}x_{\sigma(n)}]\\&&+\sum_{\sigma}(-1)^{\frac{p(p-1)}{2}+\sum_{j=1}^p\sigma(j)}N^{p}\\&&[x_{\sigma(1)},x_{\sigma(2)},\cdots,x_{\sigma(p)},N^{\alpha_{\sigma(p+1)}}x_{\sigma(p+1)},\cdots,N^{\beta_{n-1}+1}x_{n-1},N^{\alpha_{n}}x_{n}]\\&&+\sum_{\sigma}(-1)^{\frac{p(p-1)}{2}+\sum_{j=1}^{p-1}\sigma(j)+n}N^{p+\alpha_n-1}\\&&
 [x_{\sigma(1)},x_{\sigma(2)},\cdots,N^{\alpha_{n}}x_{n},N^{\alpha_{\sigma(p+1)}}x_{\sigma(p+1)},\cdots,N^{\beta_{n-1}+1}x_{n-1}]\big)\\&&+\sum_{p=1}^n\big(\sum_{\sigma}(-1)^{\frac{p(p-1)}{2}+\sum_{j=1}^{p-1}\sigma(j)+n-1}N^{p+\beta_{n-1}}\\&&[x_{\sigma(1)},x_{\sigma(2)},\cdots,N^{\beta_{n-1}}x_{n-1},N^{\alpha_{\sigma(p+1)}}x_{\sigma(p+1)},\cdots,
 N^{\alpha_{n}}x_{n}]\\
  &&+\sum_{\sigma}(-1)^{\frac{p(p-1)}{2}+\sum_{j=1}^{p-2}\sigma(j)+2n-1}N^{p+\alpha_n+\beta_{n-1}-1}\\&&[x_{\sigma(1)},x_{\sigma(2)},\cdots,x_{n-1},x_{n},N^{\alpha_{\sigma(p+1)}}x_{\sigma(p+1)},\cdots,
 N^{\alpha_{\sigma(n)}}x_{\sigma(n)}]\\&&+\sum_{\sigma}(-1)^{\frac{p(p-1)}{2}+\sum_{j=1}^p\sigma(j)}N^{p}\\&&
 [x_{\sigma(1)},x_{\sigma(2)},\cdots,x_{\sigma(p)},N^{\alpha_{\sigma(p+1)}}x_{\sigma(p+1)},\cdots,N^{\beta_{n-1}+1}x_{n-1},N^{\alpha_{n}}x_{n}]
 \end{eqnarray*}
  \begin{eqnarray*}
 &&+\sum_{\sigma}(-1)^{\frac{p(p-1)}{2}+\sum_{j=1}^{p-1}\sigma(j)+n}N^{p+\alpha_n-1}\\&&
 [x_{\sigma(1)},x_{\sigma(2)},\cdots,N^{\alpha_{n}}x_{n},N^{\alpha_{\sigma(p+1)}}x_{\sigma(p+1)},\cdots,N^{\beta_{n-1}+1}x_{n-1}]\Big)\\
 &=&\sum_{p=1}^n\Big(\sum_{\sigma}(-1)^{\frac{(p-2)(p-1)}{2}+\sum_{j=1}^{p-1}\sigma(j)}N^{p}\\&&[x_{\sigma(1)},x_{\sigma(2)},\cdots,N^{\alpha_{\sigma(p-1)}}x_{\sigma(p-1)},N^{\alpha_{\sigma(p)}}x_{\sigma(p)},\cdots, N^{\beta_{n-1}}x_{n-1},N^{\alpha_{n}}x_{n}]\\&&+\sum_{\sigma}(-1)^{\frac{(p-2)(p-1)}{2}+\sum_{j=1}^{p-2}\sigma(j)+n}N^{p+\alpha_n}\\&&[x_{\sigma(1)},x_{\sigma(2)},\cdots,x_{n},N^{\alpha_{\sigma(p)}}x_{\sigma(p)},\cdots, N^{\beta_{n-1}}x_{n-1}]\\&&+\sum_{\sigma}(-1)^{\frac{p(p-1)}{2}+\sum_{j=1}^{p-1}\sigma(j)+n-1}N^{p+\beta_{n-1}}\\&&[x_{\sigma(1)},x_{\sigma(2)},\cdots,N^{\beta_{n-1}}x_{n-1},N^{\alpha_{\sigma(p+1)}}x_{\sigma(p+1)},\cdots,
 N^{\alpha_{n}}x_{n}]\\&&+\sum_{\sigma}(-1)^{\frac{p(p-1)}{2}+\sum_{j=1}^{p-2}\sigma(j)+2n-1}N^{p+\alpha_n+\beta_{n-1}-1}\\&&[x_{\sigma(1)},x_{\sigma(2)},\cdots,x_{n-1},x_{n},N^{\alpha_{\sigma(p+1)}}x_{\sigma(p+1)},\cdots,
 N^{\alpha_{\sigma(n)}}x_{\sigma(n)}]\Big)\\
  &=&N\Big([Nx_1,Nx_2,\cdots,N^{\beta_{n-1}}x_{n-1},N^{\beta_n}x_n]+\sum_{p=1}^n\sum_{\sigma}(-1)^{\frac{p(p-1)}{2}+\sum_{j=1}^p\sigma(j)}N^{\sum_{j=1}^p\beta_{\sigma(j)}}\\&&
 [x_{\sigma(1)},x_{\sigma(2)},\cdots,x_{\sigma(p)},N^{\beta_{\sigma(p+1)}}x_{\sigma(p+1)},\cdots,N^{\beta_{\sigma(n)}}x_{\sigma(n)}]\Big),
  \end{eqnarray*}
   where $\beta_i=1,1\leq i\leq n-2$ and $\beta_{n-1}=\alpha_{n-1},\beta_n=\alpha_n$. The conclusion is that the induction can be made with respect to $\alpha_{n-1}$, starting from the Eq.$~\eqref{eq:impotant relation}$ for  $\alpha_1=1,\alpha_2=1,\cdots,\alpha_{n-1}=1,\alpha_n$ which has been already proved above. Then the Eq.$~\eqref{eq:impotant relation}$ for $\alpha_1=1,\alpha_2=1,\cdots,\alpha_{n-1},\alpha_n$ follows immediately.}

Now we assume that Eq.$~\eqref{eq:impotant relation}$ holds for $\alpha_1=1,\alpha_2=1,\cdots,\alpha_{r-1}=1,\alpha_r=1$ and arbitrary positive integer $\alpha_{r+1},\cdots,\alpha_{n-1},\alpha_n$. Then we need to show that Eq.$~\eqref{eq:impotant relation}$ holds for $\alpha_1=1,\alpha_2=1,\cdots,\alpha_{r-1}=1$ and arbitrary positive integer $\alpha_r,\alpha_{r+1},\cdots,\alpha_{n-1},\alpha_n$. Let $\alpha_r=\beta_r+1$, applying Eq.$~\eqref{eq:impotant relation}$ to the element $N^{\beta_{r}}x_{r}$ instead of the element $x_{r}$ and let $y_1=x_1,\cdots,y_r=N^{\beta_{r}}x_{r},\cdots,y_n=x_n$. By Eq. $\eqref{eq:Nijenhuis all}$, we have
\begin{eqnarray*}
&&\sum_{p=1}^n\sum_{\sigma}(-1)^{\frac{p(p-1)}{2}+\sum_{j=1}^p\sigma(j)}N^{\sum_{j=1}^p\alpha_{\sigma(j)}}\\&&
 [x_{\sigma(1)},x_{\sigma(2)},\cdots,x_{\sigma(p)},N^{\alpha_{\sigma(p+1)}}x_{\sigma(p+1)},
 \cdots,N^{\alpha_{\sigma(n)}}x_{\sigma(n)}]\\&&+[Ny_1,Ny_2,\cdots,Ny_r,N^{\alpha_{r+1}}y_{r+1},\cdots,N^{\alpha_{n-1}}y_{n-1},N^{\alpha_n}y_n]
 \end{eqnarray*}
 \begin{eqnarray*}
 &=&\sum_{p=1}^n\Big(\sum_{\sigma,\sigma(p+s)=\alpha_r}(-1)^{\frac{p(p-1)}{2}+\sum_{j=1}^p\sigma(j)}N^{\sum_{j=1}^p\alpha_{\sigma(j)}}\\&&
 [x_{\sigma(1)},x_{\sigma(2)},\cdots,x_{\sigma(p)},N^{\alpha_{\sigma(p+1)}}y_{\sigma(p+1)},\cdots,
N^{\alpha_{r}+1}x_{\sigma(p+s)},\cdots,N^{\alpha_{\sigma(n)}}x_{\sigma(n)}]\\
 &&+\sum_{\sigma,\sigma(s)=\alpha_r}(-1)^{\frac{p(p-1)}{2}+\sum_{j=1}^p\sigma(j)}N^{\sum_{j=1,j\neq s}^p\alpha_{\sigma(j)}+\alpha_r+1}\\&&
 [x_{\sigma(1)},\cdots,x_{\sigma(s)},\cdots,x_{\sigma(p)},N^{\alpha_{\sigma(p+1)}}x_{\sigma(p+1)},\cdots,N^{\alpha_{\sigma(n)}}x_{\sigma(n)}]\\
 &&-\sum_{\sigma,\sigma(p+s)=\alpha_r}(-1)^{\frac{p(p-1)}{2}+\sum_{j=1}^p\sigma(j)}N^{\sum_{j=1}^p\alpha_{\sigma(j)}}
 \\&&[x_{\sigma(1)},x_{\sigma(2)},\cdots,x_{\sigma(p)},N^{\alpha_{\sigma(p+1)}}y_{\sigma(p+1)},\cdots,N^{\alpha_{r}+1}x_{\sigma(p+s)},\cdots,N^{\alpha_{\sigma(n)}}x_{\sigma(n)}]\\
  &&-\sum_{\sigma,\sigma(s)=\alpha_r}(-1)^{\frac{p(p-1)}{2}+\sum_{j=1}^p\sigma(j)}N^{\sum_{j=1,j\neq s}^p\alpha_{\sigma(j)}+1}\\&&
 [x_{\sigma(1)},\cdots,N^{\alpha_r}x_{\sigma(s)},\cdots,x_{\sigma(p)},N^{\alpha_{\sigma(p+1)}}x_{\sigma(p+1)},\cdots,N^{\alpha_{\sigma(n)}}x_{\sigma(n)}]\Big)\\
  &=&\sum_{p=1}^n\Big(\sum_{\sigma,\sigma(s)=\alpha_r}(-1)^{\frac{p(p-1)}{2}+\sum_{j=1}^p\sigma(j)}N^{\sum_{j=1,j\neq s}^p\alpha_{\sigma(j)}+\alpha_r+1}\\&&
 [x_{\sigma(1)},\cdots,x_{\sigma(s)},\cdots,x_{\sigma(p)},N^{\alpha_{\sigma(p+1)}}x_{\sigma(p+1)},\cdots,N^{\alpha_{\sigma(n)}}x_{\sigma(n)}]\\
 &&+\sum_{\sigma,\sigma(s)=\alpha_r}(-1)^{\frac{(p-1)(p-2)}{2}+\sum_{j=1,j\neq s}^p\sigma(j)}N^{\sum_{j=1,j\neq s}^p\alpha_{\sigma(j)}+1}
 \\&&[x_{\sigma(1)},\cdots,x_{\sigma(p)},N^{\alpha_{\sigma(p+1)}}x_{\sigma(p+1)},\cdots,\stackrel{p+\alpha_r-s}{N^{\alpha_r}x_{\sigma(s)}},\cdots,N^{\alpha_{\sigma(n)}}x_{\sigma(n)}]
 \Big)\\
 &=&N\big(\sum_{p=1}^n\sum_{\sigma}(-1)^{\frac{p(p-1)}{2}+\sum_{j=1}^p\sigma(j)}N^{\sum_{j=1}^p\beta_{\sigma(j)}}\\&&
 [x_{\sigma(1)},x_{\sigma(2)},\cdots,x_{\sigma(p)},N^{\beta_{\sigma(p+1)}}x_{\sigma(p+1)},\cdots,N^{\beta_{\sigma(n)}}x_{\sigma(n)}]\\&&+[Nx_1,Nx_2,\cdots,N^{\beta_r}x_r,N^{\beta_{r+1}}y_{r+1},\cdots,N^{\beta_{n-1}}y_{n-1},N^{\beta_n}y_n]\big),
  \end{eqnarray*}
  where $\beta_i=1,1\leq i\leq r-1$ and $\beta_i=\alpha_i,r\leq i\leq n$. Therefore, Eq.$~\eqref{eq:impotant relation}$ holds for $\alpha_1=1,\alpha_2=1,\cdots,\alpha_{r-1}=1$ and arbitrary positive integer $\alpha_r,\alpha_{r+1},\cdots,\alpha_{n-1},\alpha_n$. In particular, Eq.$~\eqref{eq:impotant relation}$ holds for arbitrary positive $\alpha_1,\alpha_2,\cdots,\alpha_{n-1},\alpha_n$.

Suppose that $N$ is invertible. Applying $N^{\alpha_n}$ to Eq.$~\eqref{eq:impotant relation}$, substituting $x'_n=N^{\alpha_n}x_n$, we get

\begin{eqnarray*}
&&\sum_{p=0}^n\sum_{\sigma}(-1)^{\frac{p(p-1)}{2}+\sum_{j=1}^p\sigma(j)}N^{\sum_{j=1}^p\alpha_{\sigma(j)}-\alpha_n}\\&&[x_{\sigma(1)},x_{\sigma(2)},\cdots,x_{\sigma(p)},N^{\alpha_{\sigma(p+1)}}x_{\sigma(p+1)},\cdots,N^{\alpha_{\sigma(n)}}x_{\sigma(n)}]
\\
&=&\sum_{p=0}^n\big(\sum_{\sigma}(-1)^{\frac{p(p-1)}{2}+\sum_{j=1}^p\sigma(j)}N^{\sum_{j=1}^p\alpha_{\sigma(j)}-\alpha_n}\\&&[x_{\sigma(1)},x_{\sigma(2)},\cdots,x_{\sigma(p)},N^{\alpha_{\sigma(p+1)}}x_{\sigma(p+1)},\cdots,x'_n]\end{eqnarray*}
\begin{eqnarray*}&&
+\sum_{\sigma}(-1)^{\frac{p(p-1)}{2}+\sum_{j=1}^{p-1}\sigma(j)+n}N^{\sum_{j=1}^{p-1}\alpha_{\sigma(j)}}\\&&[x_{\sigma(1)},x_{\sigma(2)},\cdots,N^{-\alpha_n}x'_{n},N^{\alpha_{\sigma(p+1)}}x_{\sigma(p+1)},\cdots,N^{\alpha_{\sigma(n)}}x_{\sigma(n)}]\big)
\\
 &=&-\sum_{p=0}^n\big(\sum_{\sigma}(-1)^{\frac{p(p+1)}{2}+\sum_{j=1}^p\sigma(j)+n}N^{\sum_{j=1}^p\alpha_{\sigma(j)}-\alpha_n}\\&&[x_{\sigma(1)},x_{\sigma(2)},\cdots,x_{\sigma(p)},x'_n,N^{\alpha_{\sigma(p+1)}}x_{\sigma(p+1)},\cdots,N^{\alpha_{\sigma(n)}}x_{\sigma(n)}]\\&&
+\sum_{\sigma}(-1)^{\frac{(p-2)(p-1)}{2}+\sum_{j=1}^{p-1}\sigma(j)}N^{\sum_{j=1}^{p-1}\alpha_{\sigma(j)}}\\&&[x_{\sigma(1)},x_{\sigma(2)},\cdots,x_{\sigma(p-1)}N^{\alpha_{\sigma(p+1)}}x_{\sigma(p+1)},\cdots,N^{-\alpha_n}x'_{n}]\big)\\
&=&\sum_{p=0}^n\sum_{\sigma}(-1)^{\frac{p(p-1)}{2}+\sum_{j=1}^p\sigma(j)}N^{\sum_{j=1}^p\beta_{\sigma(j)}}\\&&[x_{\sigma(1)},x_{\sigma(2)},\cdots,x_{\sigma(p)},N^{\beta_{\sigma(p+1)}}x_{\sigma(p+1)},\cdots,N^{\beta_{\sigma(n)}}x_{\sigma(n)}]=0,
 \end{eqnarray*}
 where $\beta_i=\alpha_i,1\leq i\leq n-1$ and $\beta_n=-\alpha_n$.
Then the formula $\eqref{eq:impotant relation}$ holds for $\alpha_i>0,1\leq i\leq n-1$ and $\alpha_n<0$. Similarly,  the formula $\eqref{eq:impotant relation}$ holds for $\alpha_i>0,1\leq i\leq n,i\neq j$ and $\alpha_j<0$. To prove Eq.~$\eqref{eq:impotant relation}$, for $\alpha_{i_1}<0,\alpha_{i_2}<0,\cdots,\alpha_{i_r}<0$ and others positive, apply $N^{-\sum_{j=1}^r\alpha_{i_j}}$ to Eq.~$\eqref{eq:impotant relation}$ putting $x'_{i_j}=N^{\alpha_{i_j}}x_{i_j},1\leq j\leq r$. This ends the proof.\qed

\end{document}